\documentclass[pra,aps,amsmath,amssymb,amsfonts,twocolumn,nofootinbib,longbibliography,superscriptaddress]{revtex4-2}
\usepackage{amssymb}
\usepackage{bm,mathrsfs}
\usepackage{graphicx}
\usepackage{epsfig}
\usepackage{amsmath,bbm}
\usepackage{amsfonts,amssymb}
\usepackage{times}
\usepackage{verbatim}
\usepackage[sort&compress]{natbib}
\usepackage{amsmath}
\usepackage{bm}
\usepackage{float}
\usepackage{braket}
\usepackage{booktabs}
\usepackage[colorlinks,breaklinks,linkcolor=blue,anchorcolor=blue,citecolor=blue,urlcolor=blue]{hyperref}

\begin{document}
\title{Multipartite controlled-NOT gates using molecules and Rydberg atoms}
\author{Yi-Han Bai}
\affiliation{Center for Quantum Science and School of Physics, Northeast Normal University, Changchun 130024, China}
\author{Yue Wei}
 \affiliation{Center for Quantum Science and School of Physics, Northeast Normal University, Changchun 130024, China}%

\author{Chi Zhang}
 \affiliation{ 
Centre for Cold Matter, Blackett Laboratory, Imperial College London,
Prince Consort Road, London SW7 2AZ, United Kingdom}%

\author{Weibin Li}
\affiliation{School of Physics and Astronomy, and Centre for the Mathematics and Theoretical Physics of Quantum Non-equilibrium Systems, The University of Nottingham, Nottingham NG7 2RD, United Kingdom
}%

\author{Xiao-Qiang Shao}
\email[]{xqshao@nenu.edu.cn}
\affiliation{Center for Quantum Science and School of Physics, Northeast Normal University, Changchun 130024, China
}%
\affiliation{Institute of Quantum Science and Technology, Yanbian University, Yanji 133002, China}

\date{\today}

\begin{abstract}
We propose high-fidelity controlled-NOT (CNOT) gates in a hybrid system of polar molecules and Rydberg atoms based on the unconventional Rydberg pumping mechanism. By combining the rich internal structure of polar molecules with the strong dipole–dipole interactions of Rydberg atoms, we realize both two-to-one and one-to-two gate configurations. Numerical simulations show that the gate performance is robust against spontaneous emission from Rydberg states. The approach naturally extends to larger systems, as demonstrated by four-qubit implementations achieving three-to-one and one-to-three CNOT gates with fidelities exceeding 99\%. These results highlight hybrid molecule–Rydberg atom architectures as a promising platform for scalable quantum information processing.
\end{abstract}
\maketitle

Quantum computing offers the potential to solve certain problems beyond the reach of classical methods~\cite{Nielsen_Chuang_2010, PhysRevLett.79.325, PhysRevA.72.022329, PhysRevLett.95.140501}. Universal quantum computation can be achieved using single- and two-qubit gates~\cite{PhysRevA.51.1015, PhysRevA.52.3457}. However, such decompositions typically lead to deep circuits and substantial resource overhead in large-scale implementations.
Multi-qubit gates provide a direct route to reducing circuit depth and improving operational efficiency, playing a key role in quantum algorithms~\cite{Vandersypen2001,  PhysRevA.74.052318,PhysRevA.80.012312,e24101371,Zhu2025}, quantum circuits~\cite{PhysRevA.71.012337, PhysRevA.75.022313,PhysRevA.76.054301, PhysRevA.82.062326, PhysRevA.93.022311}, and quantum error correction~\cite{PhysRevLett.81.2152, PhysRevLett.86.5811, Chiaverini2004, PhysRevA.71.052332, Aoki2009}. Consequently, the realization of fast, high-fidelity, and scalable multi-qubit gates is a central challenge in quantum information processing.
Significant progress has been made across various physical platforms, including superconducting circuits~\cite{ Fedorov2012,Reed2012,PhysRevB.85.054504, PhysRevLett.114.200502,PhysRevApplied.6.054005, PhysRevApplied.14.014072, Xiao-LeLi:51205}, trapped ions~\cite{PhysRevLett.86.3907, PhysRevLett.102.040501,PhysRevA.84.012314, Figgatt2017}, photonic systems~\cite{PhysRevLett.62.2124,PhysRevA.78.032317,Lanyon2009,PhysRevA.80.042310,PhysRevLett.111.160407,Liu_2020,Li2022,https://doi.org/10.1002/qute.202400418,Toozandehjani2025}, quantum dots~\cite{PhysRevB.100.085419, Takeda2022,PhysRevA.111.042616}, and neutral atoms~\cite{PhysRevA.72.032333,Isenhower2011, PhysRevApplied.9.051001, PhysRevA.98.062338, PhysRevA.98.042704,PhysRevLett.123.170503, Yin:20, Yin20212541,WANG2025112812}. Despite these advances, achieving simultaneously high fidelity, robustness, and scalability remains an open problem. However, the implementation of multi-qubit gates is challenging in many physical systems due to operational complexity and limited coherence times. Hybrid quantum systems offer a promising route to overcoming these limitations by combining the complementary advantages of different platforms~\cite{Wallquist:2009mwr,annurev:/content/journals/10.1146/annurev-conmatphys-062910-140514, RevModPhys.85.623, annurev:/content/journals/10.1146/annurev-conmatphys-030212-184253,doi:10.1073/pnas.1419326112}. In particular, the ability to independently control distinct subsystems reduces crosstalk and improves performance in mid-circuit operations and quantum error correction. Hybrid quantum gates can generally be categorized into direct-coupling schemes, which rely on intrinsic interactions between subsystems, and mediated schemes, where interactions are established via an auxiliary system.

Neutral atoms trapped in optical tweezer arrays provide a particularly attractive platform for quantum information processing. Their high degree of reconfigurability allows flexible control of qubit connectivity, which is essential for quantum error correction and complex circuit implementations. Rydberg atoms, with their long-lived excited states~\cite{PhysRevA.79.052504} and large electric dipole moments, support strong and tunable  dipole–dipole or van der Waals interactions, including with other dipolar systems such as molecules~\cite{Flannery_2005, Browaeys_2016, Saffman_2016}. Experiments have demonstrated coherent control of large-scale Rydberg atom arrays with long coherence times~\cite{10.1093/nsr/nww013,doi:10.1126/science.aal3837, Picken_2019}, establishing them as a powerful platform for hybrid quantum gate implementations.
Hybrid architectures involving Rydberg atoms have been explored in several directions. 
Ion–atom systems combine the scalability of neutral atoms with the precise addressability of trapped ions to realize high-fidelity quantum gates~\cite{PhysRevA.76.033409, PhysRevA.81.012708, PhysRevA.94.013420,RevModPhys.91.035001, Shaposhnikov2023}.
Atom–photon systems exploit strong light–matter interactions in cavity quantum electrodynamics for quantum gate implementation~\cite{PhysRevLett.92.127902, doi:10.1126/science.1246164, Reiserer2014, PhysRevLett.126.130502, PhysRevA.109.032602}. Specifically, molecule–atom hybrid systems provide a compelling approach: polar molecules offer rich internal structure~\cite{Sawant_2020}, long coherence times~\cite{Gregory2021, PhysRevLett.127.123202}, and electric dipole moments in their electronic ground states~\cite{Yan2013, doi:10.1126/science.abn8525}. Leveraging dipole–dipole interactions between molecules and Rydberg atoms, recent works have demonstrated fast and high-fidelity Rydberg-mediated molecular quantum gates~\cite{PRXQuantum.3.030339, PRXQuantum.3.030340, Zhang2026HybridGate}.

In this paper, we propose two types of CNOT gates in a hybrid molecule–atom system: many-to-one and one-to-many configurations, implemented via the unconventional Rydberg pumping (URP) mechanism~\cite{PhysRevA.98.062338}. The scheme exploits the complementary roles of the two subsystems: polar molecules provide stable, interacting ground states with long coherence times, while Rydberg atoms enable strong and tunable interactions for fast state transfer.
A key advantage of this approach is that molecular qubits remain in their electronic ground states during the gate operation, where dipole interactions allow trapping and suppress motional decoherence and loss. Meanwhile, Rydberg excitation facilitates efficient interactions, leading to high-fidelity gate operations with enhanced robustness. These features make the scheme well suited for implementing complex quantum operations in hybrid architectures.

\begin{figure*}
\centering
\includegraphics[width=0.8\linewidth]{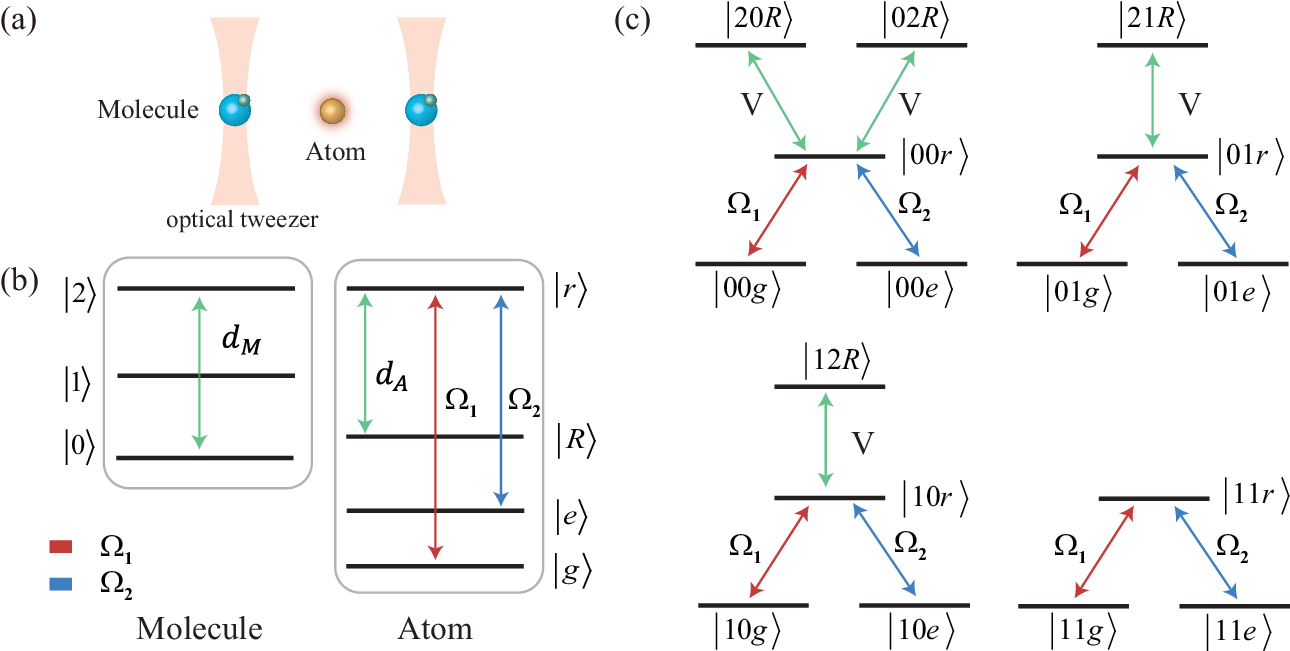}
\caption{ System configuration and operational mechanism of the hybrid many-to-one CNOT gate. (a) Spatial arrangement of the $^{87}\mathrm{Rb}$ atom and $\mathrm{CaF}$ molecules. Two molecules (control qubits, left and right) and a central Rydberg atom (target qubit) are aligned linearly, with the atom positioned equidistant from the molecules. (b) Energy-level structure of the molecule–atom system. The molecule contains three states $|0\rangle$, $|1\rangle$, and $|2\rangle$. The atom involves two ground states $|g\rangle$, $|e\rangle$, and two Rydberg states $|R\rangle$, $|r\rangle$. The green lines represent the transition dipole moments that mediate the interaction. The red and blue arrows denote the laser fields with Rabi frequencies $\Omega_1$ and $\Omega_2$, respectively, which couple the atomic ground states to the Rydberg state $|r\rangle$. (c) Transition pathways in different subspaces. $V$ denotes the strength of the molecule–atom dipole–dipole interaction. }
\label{Figure1}
\end{figure*}

For the hybrid many-to-one CNOT gate, we consider two $\mathrm{CaF}$ molecular control qubits and one $^{87}\mathrm{Rb}$ atomic target qubit arranged in a symmetric geometry, with the atom located equidistant from the two molecules, as shown in Fig.~\ref{Figure1}(a). Although this symmetric configuration is adopted for clarity, identical interaction strengths are not required, and the gate mechanism remains robust against moderate variations in the relative positions of the particles.

Neglecting the much weaker molecule–molecule interaction~\cite{PhysRevLett.131.013401,PhysRevLett.134.133401}, the dynamics are dominated by the molecule–atom dipole–dipole interaction. The corresponding energy-level structure is illustrated in Fig.~\ref{Figure1}(b). Each molecular control qubit is described by three states $|0\rangle$, $|1\rangle$, and $|2\rangle$, where the transition $|0\rangle \leftrightarrow |2\rangle$ carries a dipole moment $d_M$. In contrast, the atomic target qubit comprises two ground states $|g\rangle$ and $|e\rangle$, together with two Rydberg states $|R\rangle$ and $|r\rangle$, with a dipole moment $d_A$ associated with the transition $|R\rangle \leftrightarrow |r\rangle$.
Resonant laser fields couple the Rydberg state $|r\rangle$ to $|g\rangle$ and $|e\rangle$ with Rabi frequencies $\Omega_1$ and $\Omega_2$. At the same time, tuning the atomic transition $|R\rangle \leftrightarrow |r\rangle$ into resonance with the molecular transition $|0\rangle \leftrightarrow |2\rangle$ induces a molecule–atom dipole–dipole interaction through the coupling of $d_M$ and $d_A$ (see the appendix for details). The total Hamiltonian of the system can then be written as ($\hbar = 1$)
\begin{equation}\label{eq:1}
\hat{H}_1 =
\frac{\Omega_{1}}{2} |g\rangle \langle r|
+\frac{\Omega_{2}}{2} |e\rangle \langle r|
+\frac{V}{2}\sum_{j=1}^{2} |0r\rangle_j \langle 2R|
+ \mathrm{H.c.},
\end{equation}
where $V$ denotes the strength of the molecule–atom dipole–dipole interaction, and $j$ labels the two molecular qubits.

As illustrated in Fig.~\ref{Figure1}(c), transition pathways are analyzed within subspaces spanned by $\{|00g\rangle, |00e\rangle, |01g\rangle, |01e\rangle, |10g\rangle, |10e\rangle, |11g\rangle, |11e\rangle\}$. The URP regime is considered with molecule–atom dipole–dipole interaction satisfying $V \gg \{\Omega_1,\Omega_2\}$.

When at least one molecule occupies $|0\rangle$ (i.e., in $|00\rangle$, $|01\rangle$, and $|10\rangle$ subspaces), the atomic Rydberg state $|r\rangle$ couples resonantly to the molecular auxiliary state through $|0r\rangle \leftrightarrow |2R\rangle$. Strong coupling hybridizes these states into dressed states separated by an energy on the order of $V$. This Autler–Townes splitting shifts the excitation out of resonance and suppresses laser-driven population transfer to the Rydberg manifold. Consequently, transitions from $|g\rangle$ and $|e\rangle$ to $|r\rangle$ are strongly inhibited, leaving the atomic state unchanged (see the appendix for details).

By contrast, in $|11g\rangle$ and $|11e\rangle$, the dipole–dipole interaction vanishes. Dynamics are then confined to the subspace spanned by $|11g\rangle$, $|11e\rangle$, and $|11r\rangle$, as shown in Fig.~\ref{Figure1}(c). Under condition $V \gg {\Omega_1,\Omega_2}$, the effective Hamiltonian reduces to
\begin{equation}
\hat{H}^\mathrm{eff}_{1}
=\frac{\Omega_{1}}{2}|11g\rangle \langle 11r|+ \frac{\Omega_{2}}{2}|11e\rangle \langle 11r|
+ \mathrm{H.c.}
\label{eq:2}
\end{equation}

Gate performance is quantified by fidelity $F = |\langle \Psi_{\text{ideal}}|\Psi_t\rangle|^2$,
where $|\Psi_{\text{ideal}}\rangle$ denotes the ideal output state and $|\Psi_t\rangle$ represents the system state at time $t$. To enhance both fidelity and robustness, a time-dependent Gaussian pulse is applied,
$\Omega(t) = \Omega_{\text{max}} \exp[{-{(t-t_0)^2}/{2\sigma^2}}],$
with $\Omega_{\text{max}}$ and $\sigma$ denoting peak amplitude and pulse width, respectively.
Setting $\Omega_1(t)=-\Omega_2(t)=\Omega(t)$ gives an effective Rabi frequency $\Omega_{\mathrm{eff}}(t)=\sqrt{\Omega_1^2+\Omega_2^2}=\sqrt{2}\Omega(t)$. Gate operation follows from the pulse-area condition $\int_0^{T} \Omega_{\mathrm{eff}}(t) dt = 2\pi$.

\begin{figure}
\centering
\includegraphics[width=1\linewidth]{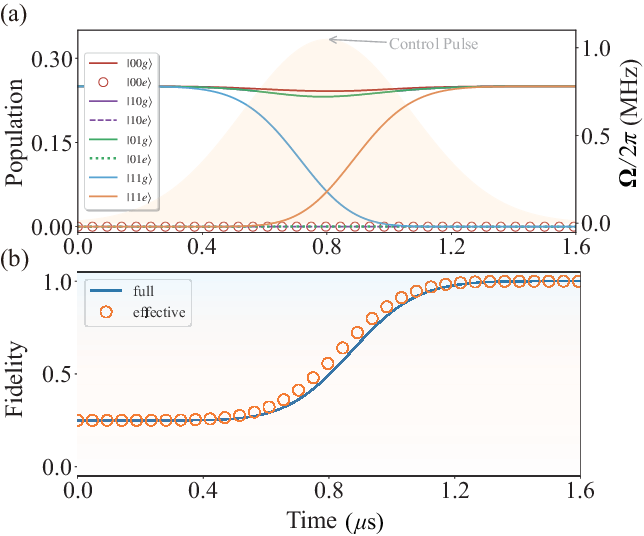}
\caption{Dynamics of the hybrid two-to-one CNOT gate driven by Gaussian pulses. (a) The populations of the relevant basis states and the control pulse. (b) Time evolution of the fidelity. The blue solid line and the orange open circles represent the dynamics governed by the full Hamiltonian and the effective Hamiltonian, respectively. The system is driven by Gaussian pulses defined as
$\Omega_1=-\Omega_2=\Omega=\Omega_{\text{max}} \exp[^{-(t-t_0)^2/(2\sigma^2)}]$,
where the maximum Rabi frequency is $\Omega_{\text{max}} = 2\pi \times 1.05$ MHz and the pulse width is $\sigma = 0.27 \mu$s. The molecule–atom interaction strength is set to $V = 2\pi \times 4.04$ MHz.}
\label{Figure2}
\end{figure}

Figure~\ref{Figure2}(a) shows population dynamics together with a Gaussian pulse. The atomic qubit flips between $|g\rangle$ and $|e\rangle$ only when both molecular qubits occupy $|1\rangle$. If either molecule is in $|0\rangle$, the strong dipole–dipole interaction suppresses excitation, leaving the atomic state unchanged. Figure~\ref{Figure2}(b) compares fidelity dynamics obtained from the full Hamiltonian $\hat{H}_1$ in Eq.~(\ref{eq:1}) and the effective Hamiltonian $\hat{H}^{\mathrm{eff}}_1$ in Eq.~(\ref{eq:2}). Excellent agreement, with fidelity approaching unity at optimal gate time, confirms the validity of the effective model and demonstrates the high performance of the proposed scheme.
\begin{figure*}
\centering
\includegraphics[width=0.8\linewidth]{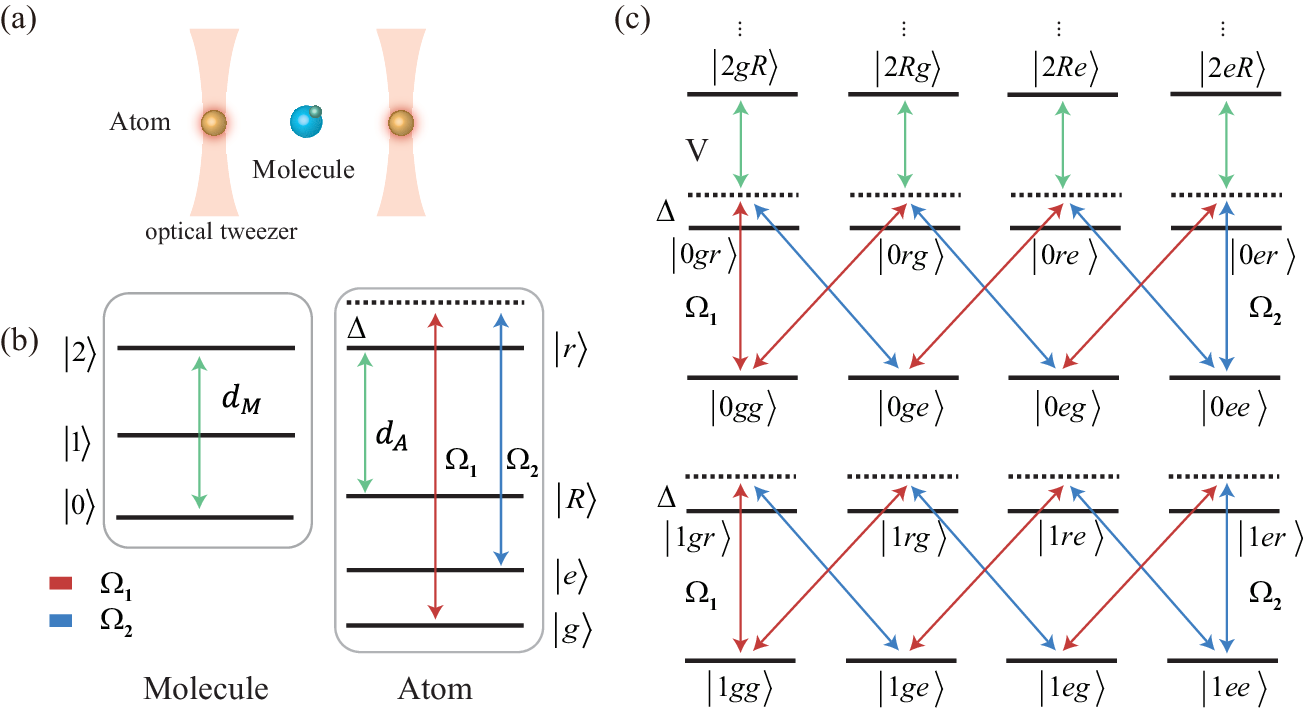}
\caption{ System configuration and operating mechanism of the hybrid one-to-many CNOT gate. (a) Spatial arrangement where two Rydberg atoms (target qubits) are positioned symmetrically on either side of a central molecule (control qubit). (b) Relevant energy-level structures. The molecular energy levels are the same as those shown in Fig.~\ref{Figure1}(b). For the atoms, two laser fields with Rabi frequencies $\Omega_1$ (red arrow) and $\Omega_2$ (blue arrow) drive transitions from the ground states to the Rydberg state $|r\rangle$ with a detuning $\Delta$. 
(c) Transition pathways in different subspaces. The upper panels correspond to the case where the control molecule is in state $|0\rangle$, while the lower panels correspond to the case where the molecule is in state $|1\rangle$.}
\label{Figure3}
\end{figure*}

Having established the synthesis methodology for two-to-one operations—which primarily addresses the challenge of multi-conditional logic—we now shift our focus to its architectural counterpart. In many quantum algorithms, it is equally crucial to fan out a single conditional trigger to multiple qubits.
The spatial configuration is shown in Fig.~\ref{Figure3}(a), where the molecule is located equidistant from the two Rydberg atoms. In this scheme, each atom is driven by two laser fields with Rabi frequencies $\Omega_1$ and $\Omega_2$, coupling the transitions $|g\rangle \leftrightarrow |r\rangle$ and $|e\rangle \leftrightarrow |r\rangle$, respectively, with a common detuning $\Delta$, as illustrated in Fig.~\ref{Figure3}(b). The Hamiltonian of the hybrid system is given by
\begin{eqnarray}
\hat H_{2} &= & \frac{\Omega_{1}}{2}\sum_{k=1}^{2} |g\rangle_k \langle r|
+ \frac{\Omega_{2}}{2}\sum_{k=1}^{2} |e\rangle_k \langle r|\notag \nonumber\\&&
+ \frac{V}{2}\sum_{k=1}^{2} |0r\rangle_k \langle 2R|
+ \mathrm{H.c.} \notag \nonumber\\&& + U_1 |rr\rangle \langle rr|
+ U_2 |RR\rangle \langle RR|
- \Delta \sum_{k=1}^{2} |r\rangle_k \langle r| ,
\label{eq:3}
\end{eqnarray}
where $U_1$ and $U_2$ denote van der Waals interaction strengths, and $k$ labels the two atomic qubits.

To simplify the dynamics, the Hamiltonian $\hat{H}_2$ is transformed into a rotating frame defined by
$U = \exp\left[-it\left(U_{1} |rr\rangle \langle rr| + U_{2} |RR\rangle \langle RR|\right)\right]$.
The transformed Hamiltonian reads
$\hat H_2' = U^\dagger \hat H_2 U + i\dot U^\dagger U$.
Under strong interactions $\{U_1 , U_2\}\gg\{\Delta, V, \Omega_1, \Omega_2\}$, rapidly oscillating terms associated with Rydberg--Rydberg interactions can be neglected. The analysis is  restricted to the computational subspace spanned by $
\{|0gg\rangle, |0ge\rangle, |0eg\rangle, |0ee\rangle, |1gg\rangle, |1ge\rangle, |1eg\rangle, |1ee\rangle\}$.

As shown in the upper panels of Fig.~\ref{Figure3}(c), when the molecular control qubit is in state $|0\rangle$, molecule-atom interaction dominates the dynamics. In the URP regime $\{V, \Delta\} \gg \{\Omega_{1}, \Omega_{2}\}$, intermediate states involving Rydberg excitation (e.g., $|0gr\rangle$) are resonantly coupled to states such as $|2gR\rangle$. This coupling generates dressed states with large energy shifts, effectively moving them out of resonance with the driving fields. As a result, transitions within this subspace are strongly suppressed, and computational basis states remain unchanged when the molecular qubit is in $|0\rangle$.
When the molecular control qubit occupies state $|1\rangle$, the Hamiltonian becomes
\begin{eqnarray}
{\hat H^{1}_2}&=&\frac{\Omega_{1}}{2}({P}_{1}^{1}{S}_{2}^{1}{P}_{3}^{g}+{P}_{1}^{1}{S}_{2}^{1}{P}_{3}^{e}
+{P}_{1}^{1}{P}_{2}^{g}{S}_{3}^{1}\nonumber\\&&
+{P}_{1}^{1}{P}_{2}^{e}{S}_{3}^{1})
+\frac{\Omega_{2}}{2}({P}_{1}^{1}{S}_{2}^{2}{P}_{3}^{g}+{P}_{1}^{1}{S}_{2}^{2}{P}_{3}^{e}\nonumber\\&&
+{P}_{1}^{1}{P}_{2}^{g}{S}_{3}^{2}
+{P}_{1}^{1}{P}_{2}^{e}{S}_{3}^{2})
+\mathrm{H.c.}\nonumber\\&&
-\Delta({P}_{1}^{1}{P}_{2}^{g}{P}_{3}^{r}+{P}_{1}^{1}{P}_{2}^{e}{P}_{3}^{r}+{P}_{1}^{1}{P}_{2}^{r}{P}_{3}^{g}\nonumber\\&&+{P}_{1}^{1}{P}_{2}^{r}{P}_{3}^{e}),
\label{eq:6}
\end{eqnarray}
where operators are defined as $P_n^k =|k\rangle_n\langle k|$ (for $k \in \{0,1,2\}$ or $\{g,e,r,R\}$) and $\sigma _{n}=|0\rangle _{n}\langle  2|$. Atomic operators are given by ${S} _{n}^{1}=|g\rangle _{n}\langle  r|$, ${S} _{n}^{2}=|e\rangle _{n}\langle  r|$, and ${S} _{n}^{3}=|r\rangle _{n}\langle  R|$. The index ${n}\in{\{1, 2, 3}\}$ labels qubit positions, with $n=1$ for the molecule and $n=2,3$ for the atoms (see the appendix for details).

Under the condition $\Delta \gg \{ \Omega_1, \Omega_2 \}$, the effective Hamiltonian can be derived to second order in perturbation theory as
\begin{eqnarray}
{\hat
H^{\textup{eff}}_2}&=&\frac{\Omega_{1}\Omega_{2}}{4\Delta}(|1gg\rangle+|1ee\rangle)(\langle1ge|+\langle1eg|)+\mathrm{H.c.}\nonumber\\&&+\frac{\Omega_{1}^2}{4\Delta}(2|1gg\rangle\langle1gg|+|1ge\rangle\langle1ge|+|1eg\rangle\langle1eg|)\nonumber\\&& + \frac{\Omega_{2}^2}{4\Delta}(2|1ee\rangle\langle 1ee|+|1ge\rangle\langle1ge|+|1eg\rangle\langle1eg|).
\label{eq:7}
\end{eqnarray}

Dynamics within the ground-state subspace $\{|1gg\rangle, |1ge\rangle, |1eg\rangle, |1ee\rangle\}$ are governed by effective transitions mediated by virtual excitations, as shown in the lower panels of Fig.~\ref{Figure3}(c).
Consistent with the previous scheme, a time-dependent Gaussian pulse is applied. When $\Omega_1(t)=\Omega_2(t)=\Omega(t)$, the effective Rabi frequency becomes $\Omega_{\textup{eff}}(t)=\Omega(t)^2/\Delta$. The gate operation is achieved by satisfying the pulse-area condition
$\int_0^{T} \Omega^2(t)/\Delta dt = 2\pi$.

\begin{figure}
\centering
\includegraphics[width=1\linewidth]{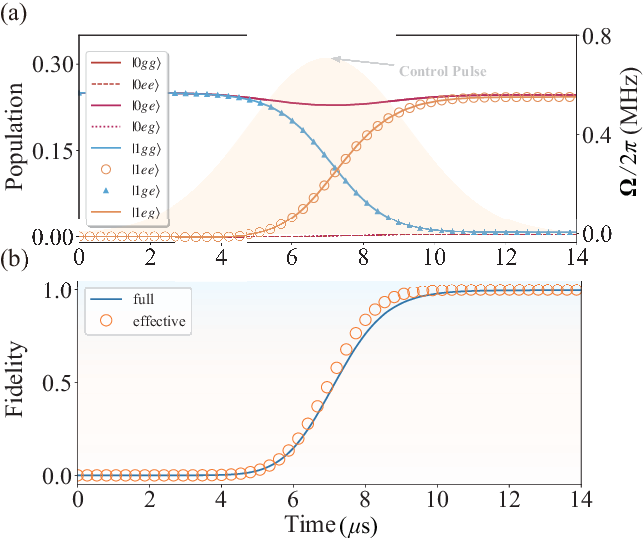}
\caption{Dynamics of the hybrid one-to-two CNOT gate driven by Gaussian pulses. (a) The populations of the relevant basis states and the control pulse. (b) Time evolution of the fidelity. The van der Waals interaction strengths are $U_1 = 2\pi \times 1774.6$ MHz and $U_2 = 2\pi \times 126.9$ MHz. The other parameters are $\Omega_{\text{max}} = 2\pi \times 0.71$ MHz, 
$\Delta = 2\pi \times 2$ MHz, and $\sigma = 2.25 \mu$s.}
\label{Figure4} 
\end{figure}

\begin{figure}
\centering
\includegraphics[width=1\linewidth]{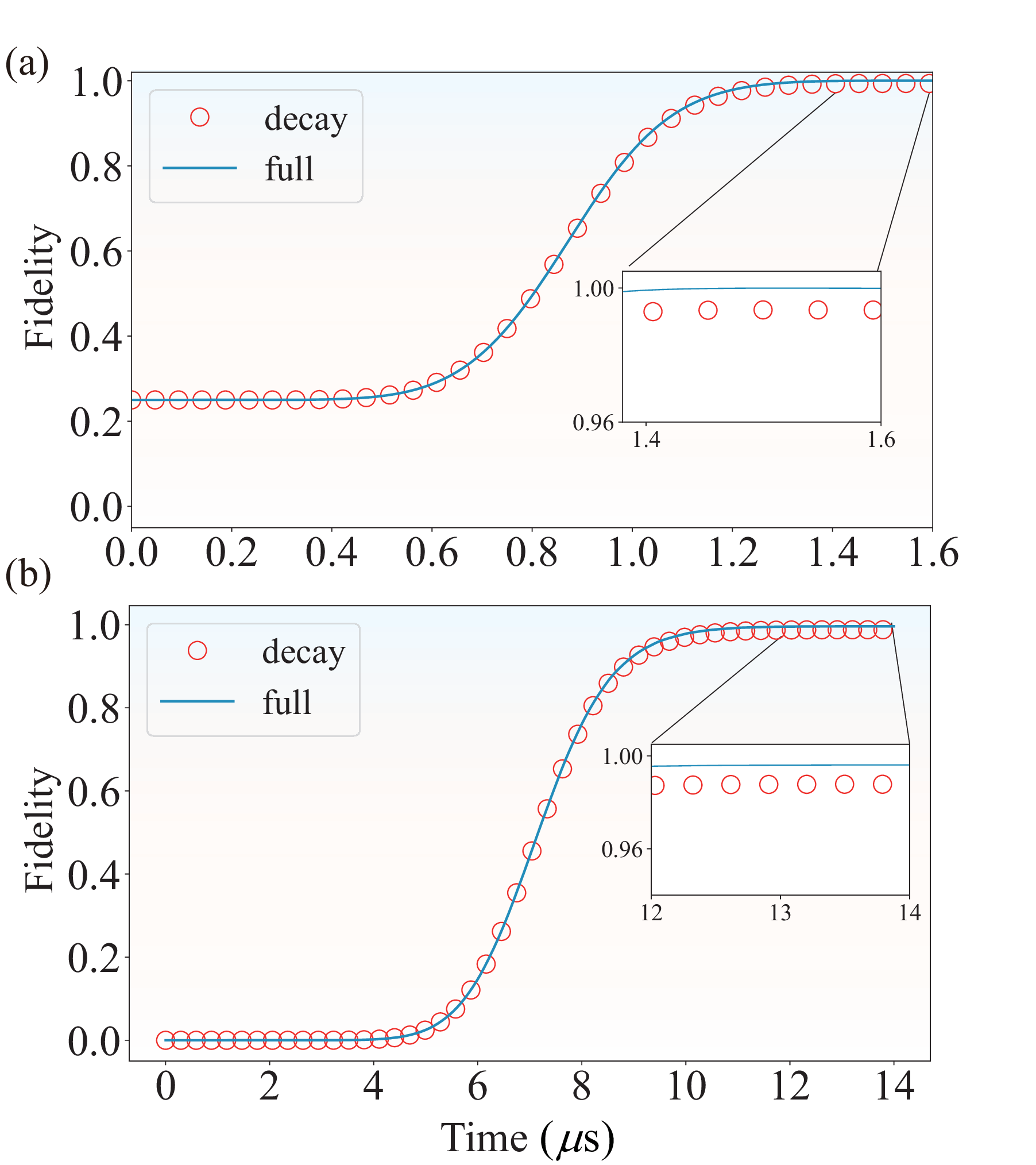}
\caption{Time evolution of the gate fidelity governed by the master equation. The blue curve corresponds to the non-dissipative case, while the red open circle includes the effect of spontaneous decay from Rydberg states. The insets show a magnified view of the fidelity overlap. The decay rates are set to 
$\gamma_1 = 2\pi \times 4.580$ kHz and $\gamma_2 = 2\pi \times 2.393$ kHz. 
(a) Hybrid two-to-one CNOT gate, with the other parameters the same as those in Fig.~\ref{Figure2}. (b) Hybrid one-to-two CNOT gate, with the other parameters the same as those in Fig.~\ref{Figure4}.}
\label{Figure5} 
\end{figure}

As illustrated in Fig.~\ref{Figure4}, the dynamics of the hybrid one-to-two CNOT gate are investigated. Results in Fig.~\ref{Figure4}(a) show that transitions $|g\rangle \leftrightarrow |e\rangle$ of the atomic qubits occur only when the molecular control qubit is in state $|1\rangle$; otherwise, the computational states remain unchanged. Figure~\ref{Figure4}(b) compares the fidelity obtained from the full Hamiltonian $\hat{H}_2$ in Eq.~(\ref{eq:3}) and the effective Hamiltonian $\hat{H}^{\text{eff}}_2$ in Eq.~(\ref{eq:7}). Excellent agreement is observed, with fidelity approaching unity at the optimal gate time. This confirms the validity of the effective Hamiltonian and demonstrates the robustness of the proposed scheme.

Although a native single-pulse multi-target gate often requires a longer operation time than a standard two-qubit CNOT, exploring its implementation is highly necessary. In practical quantum control, the overall fidelity is limited not only by decoherence but also heavily by control errors, crosstalk, and state leakage. By replacing a deep cascade of CNOTs with a single, carefully shaped driving field, we trade a slight increase in decoherence time for a massive reduction in pulse-edge errors and algorithmic circuit depth. This holistic error mitigation makes such long-duration multi-target gates particularly valuable for efficient circuit compilation and robust quantum error correction.

Under realistic experimental conditions, the effect of spontaneous decay from Rydberg states on gate performance is analyzed. Molecular qubits remain in their electronic ground states and are assumed to be free of decay. Attention is therefore restricted to the Rydberg states $|r\rangle = |59s_{1/2}, m_j = -1/2\rangle$ and $|R\rangle = |58p_{3/2}, m_j = -3/2\rangle$ of the $^{87}\mathrm{Rb}$ atom, with decay rates $\gamma_1 = 2\pi \times 4.580$ kHz and $\gamma_2 = 2\pi \times 2.393$ kHz, respectively.
Both Rydberg states are assumed to decay to the ground states $|g\rangle$ and $|e\rangle$ with equal branching ratios, resulting in decay rates of $\gamma_1/2$ and $\gamma_2/2$ for each channel~\cite{PhysRevA.98.062338, 7zjs-73qm}. Under these assumptions, the time evolution of the system density matrix is governed by the master equation
$\dot{\rho}_{i} = -i[\hat{H}_{i}, \rho_{i}] + \mathcal{L}(\rho_{i}),$
where $i=1,2$ denotes the hybrid two-to-one and one-to-two CNOT gates, respectively. $\hat{H}_{i}$ and $\rho_{i}$ denote the corresponding Hamiltonian and density matrix of each system. $\mathcal{L}$ represents the Lindblad superoperator $\mathcal{L}(\rho_{i}) = \sum_{j} \sum_{k} \left( {L}_{j}^{k} \rho_{i} {L}_{j}^{k\dagger} - \frac{1}{2} \left\{ {L}_{j}^{k\dagger} {L}_{j}^{k},\rho_{i} \right\} \right)$, where the Lindblad operators for the $j$th atom are defined as $ {L}_j^{0}=\sqrt{\gamma_1/2}|g\rangle_j\langle r|,{L}_j^{1}=\sqrt{\gamma_1/2}|e\rangle_j\langle r|, {L}_j^{2}=\sqrt{\gamma_2/2}|g\rangle_j\langle R|, {L}_j^{3}=\sqrt{\gamma_2/2}|e\rangle_j\langle R|$.
Figure~\ref{Figure5} presents the fidelity evolution for both the hybrid two-to-one [Fig.~\ref{Figure5}(a)] and one-to-two [Fig.~\ref{Figure5}(b)] CNOT gates under dissipative conditions, where the fidelity is defined as $F=\text{Tr}(\rho_\text{ideal}\rho)=\langle\Psi_{\text{ideal}}|\rho|\Psi_{\text{ideal}}\rangle$.
Blue curves show results obtained from full Hamiltonian dynamics, while red open circles correspond to dynamics governed by the master equation. As indicated in the insets, gate fidelity exhibits a slight reduction due to the spontaneous decay of Rydberg states.
Despite this effect, both hybrid systems display strong agreement between dissipative and coherent dynamics, with final fidelity remaining high. These results demonstrate that the proposed schemes are robust against the spontaneous decay of Rydberg states.
\begin{figure}
\centering
\includegraphics[width=1\linewidth]{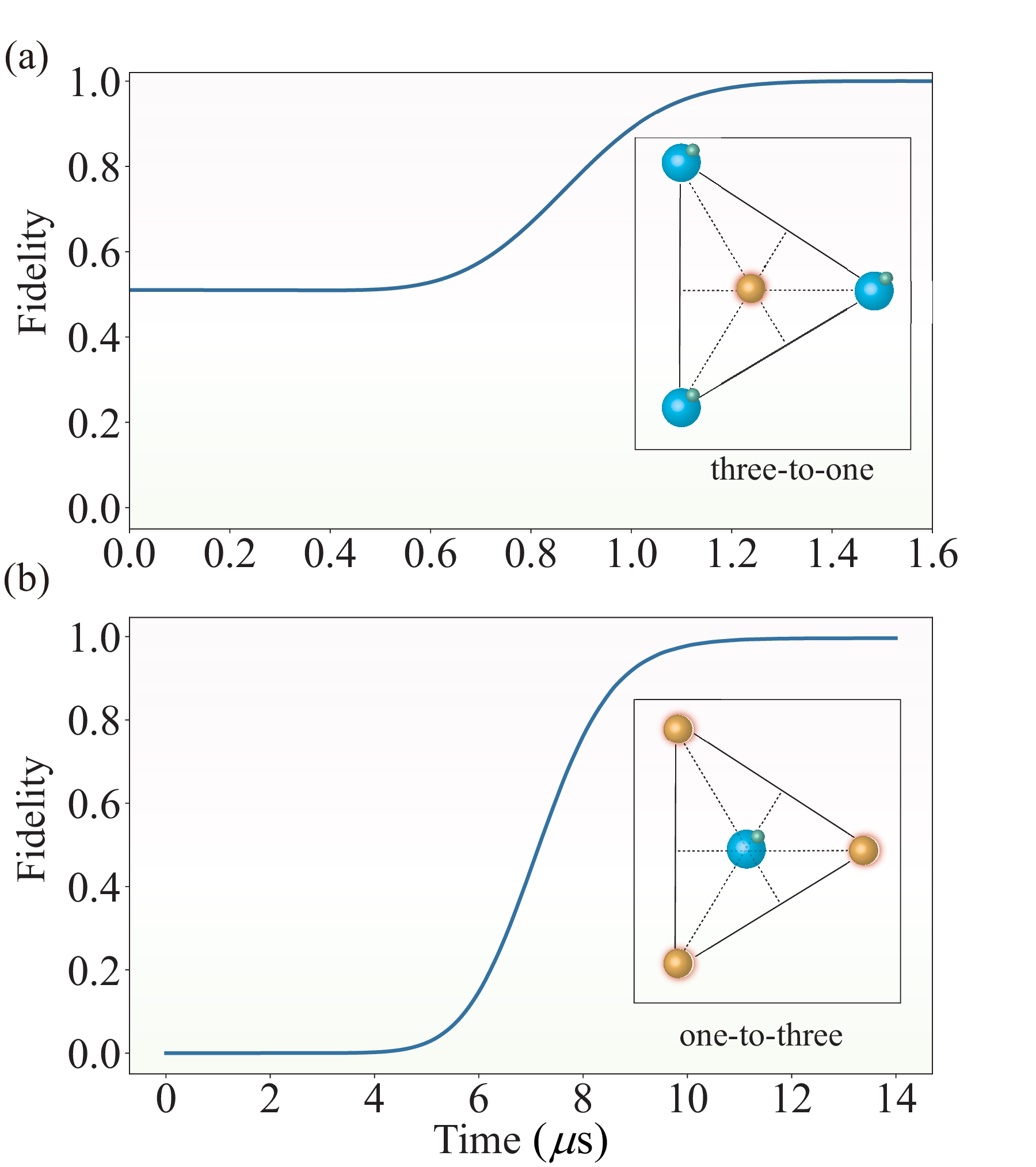}
\caption{ Scalable implementation of four-qubit CNOT gates. The molecule–atom interaction strengths are given in the appendix. (a) Spatial configuration and fidelity evolution of the three-to-one gate. The parameters are identical to Fig.~\ref{Figure2}. (b) Spatial configuration and fidelity evolution of the one-to-three gate. The parameters are identical to Fig.~\ref{Figure4}, except for the Rydberg-Rydberg interaction strengths, which are set to $U_1=2\pi\times4206.4$ MHz and $U_2=2\pi\times300.8$ MHz. }
\label{Figure6}
\end{figure}

Finally, the scalability of the proposal is demonstrated by extending the schemes to four-qubit implementations, namely three-to-one and one-to-three CNOT gates. For the three-to-one CNOT gate, the system is expanded by introducing an additional molecule, as shown in the inset of Fig.~\ref{Figure6}(a), where three molecules are arranged at the vertices of an equilateral triangle, and the target Rydberg atom is located at the centroid, such that molecule–molecule interactions remain negligible. For the one-to-three CNOT gate, an additional Rydberg atom is incorporated into the target array, as illustrated in the inset of Fig.~\ref{Figure6}(b), with interaction strengths set to $U_1=2\pi\times4206.4$ MHz and $U_2=2\pi\times300.8$ MHz to account for the modified interatomic spacing. The time evolution of fidelity for both schemes is obtained numerically, as shown in Fig.~\ref{Figure6}(a) and (b), using parameters consistent with the three-qubit cases. In both configurations, fidelity increases smoothly during gate operation and saturates above $99\%$. These results confirm that the proposed four-qubit CNOT gates can be realized with high fidelity and maintain strong robustness in larger-scale systems (see the appendix for details).

In conclusion, we propose and numerically verify a scheme for implementing high-fidelity multipartite CNOT gates in a hybrid molecule–atom system based on the URP mechanism. Simulations show strong robustness against spontaneous decay of Rydberg states. Using three- and four-qubit implementations, both many-to-one and one-to-many CNOT gates are realized, highlighting the potential of hybrid molecule–atom platforms for quantum information processing. The approach can be extended to $N$-qubit systems, offering a route to enhanced computational capability and more efficient quantum error correction. While increasing system size leads to more complex interaction structures and places stricter demands on fidelity and theoretical analysis, the present results indicate a viable path toward scalable quantum computing and the simulation of complex many-body systems.

This work is supported by the National Natural Science
Foundation of China (NSFC) under Grant No. 12174048. W.L. acknowledges support from the EPSRC through Grant No.
EP/W015641/1, and the Going Global Partnerships Programme of the British Council (Contract No. IND/CONT/G/22-23/26).

\begin{appendix}
\section{The molecule–atom dipole–dipole interaction}

The dipole–dipole interaction between the atom and the molecule arises from the electrostatic interaction between two electric dipoles. In the near-field regime, where the interparticle distance is much smaller than the radiation wavelength, the interaction Hamiltonian can be written as~\cite{Jackson1999,Saffman2010}
\begin{equation}
\hat H_{\text{DD}}=\frac{1}{4\pi\epsilon_0}
 \left(\frac{\mathbf{d}_A\cdot\mathbf{d}_M-3(\mathbf{d}_A\cdot\mathbf{n})(\mathbf{d}_M\cdot\mathbf{n})}{R^3}\right),
\end{equation}
where $\mathbf{d}_A=(d_{Ax},d_{Ay},d_{Az})$ and $\mathbf{d}_M=(d_{Mx},d_{My},d_{Mz})$ denote the electric dipole moment operators of the atom and molecule, respectively. The distance between them is $R=|\mathbf{r}_A-\mathbf{r}_M|$, and $\mathbf{n}=(\sin\theta\cos\varphi,\sin\theta\sin\varphi,\cos\theta)$ is the unit vector connecting the atom and molecule. The quantization axis is chosen along the $z$-direction, where $\theta$ denotes the angle between $\mathbf{n}$ and the $z$-axis, and $\varphi$ is the corresponding azimuthal angle.
In the spherical basis, the dipole operators can be written as
\begin{equation}
\begin{cases}
d_0=d_z,\\
d_+ = -(d_x + i d_y) / \sqrt{2}, \\
d_- = (d_x - i d_y) / \sqrt{2},
\end{cases}
\end{equation}
where $d_0$ corresponds to transitions with $\Delta m_j=0$, while $d_{\pm}$ corresponds to transitions with $\Delta m_j=\pm1$. In the spherical basis, the dipole–dipole interaction can then be written as:
\begin{equation}
\begin{aligned}
\hat{H}_{\text{DD}} &= \frac{1}{4\pi\epsilon_{0}R^{3}} \Bigg[
\frac{1 - 3 \cos^2 \theta}{2}
\left( d_{A}^{+} d_{M}^{-} + d_{A}^{-} d_{M}^{+} + 2 d_{A}^{0} d_{M}^{0} \right) \\
&\quad + \frac{3}{\sqrt{2}} \sin \theta \cos \theta
\left( d_{A}^{+} d_{M}^{0} e^{-i \varphi} - d_{A}^{-} d_{M}^{0} e^{i \varphi} \right. \\
&\qquad \left. + d_{A}^{0} d_{M}^{+} e^{-i \varphi} - d_{A}^{0} d_{M}^{-} e^{i \varphi} \right) \\
&\quad - \frac{3}{2} \sin^2 \theta
\left( d_{A}^{+} d_{M}^{+} e^{-2i \varphi} + d_{A}^{-} d_{M}^{-} e^{2i \varphi} \right)
\Bigg].
\end{aligned}
\label{HDD}
\end{equation}

To verify the experimental feasibility of the proposed scheme, it is necessary to estimate the strength of the molecule–atom dipole–dipole interaction $V$ for a realistic physical system. In this work, we consider a hybrid platform composed of $^{87}\mathrm{Rb}$ atoms and $\mathrm{CaF}$ molecules as the physical implementation~\cite{PRXQuantum.3.030340}. 
For the $^{87}\mathrm{Rb}$ atom, we select the following two Rydberg states
\begin{eqnarray}
\begin{aligned}
	|r\rangle &= |59s_{1/2}, m_j = -1/2\rangle, \\
	|R\rangle &= |58p_{3/2}, m_j = -3/2\rangle.
\end{aligned}
\end{eqnarray}
For the $\mathrm{CaF}$ molecule, we choose the rotational states
\begin{eqnarray}
\begin{aligned}
	|0\rangle &= |0,0,0\rangle ,\\
	|1\rangle &= |1,0,0\rangle ,\\
	|2_{\pm}\rangle &= |1,1,\pm1\rangle .
\end{aligned}
\end{eqnarray}
To achieve resonance between the molecular and atomic transitions, the molecular energy levels can be tuned via the Stark effect. By applying a dc electric field with a strength in the range of $1\sim2~\mathrm{V/cm}$, the molecular rotational transition $|0\rangle \leftrightarrow |2\rangle$ can be brought into resonance with the atomic Rydberg transition $|r\rangle \leftrightarrow |R\rangle$. Based on the chosen energy levels and the previously obtained transition dipole moments together with the dipole–dipole interaction formula, we estimate the interaction strength for an atom–molecule separation of $R=1~\mu\mathrm{m}$. Due to the angular momentum selection rules, the Hamiltonian $\hat{H}_{\mathrm{DD}}$ couples the state $|0r\rangle$ to two degenerate molecular auxiliary states $|2_{+}R\rangle$ and $|2_{-}R\rangle$. The corresponding coupling matrix elements are calculated as
\begin{eqnarray}
\begin{aligned}
	\frac{V_{\text{DD},+}}{2} &= \langle 0r | \hat{H}_{\text{DD}} | 2_+ R \rangle \approx -2\pi \times 0.64 \text{ MHz}, \\
	\frac{V_{\text{DD},-}}{2} &= \langle 0r | \hat{H}_{\text{DD}} | 2_- R \rangle \approx 2\pi \times 1.91 \text{ MHz}.
\end{aligned}
\end{eqnarray}
To simplify the model, we introduce an effective superposition state $|2\rangle$ defined as
\begin{equation}
|2\rangle =
\frac{1}{\sqrt{V_{\mathrm{DD},+}^{2}+V_{\mathrm{DD},-}^{2}}}
\left(
V_{\mathrm{DD},+}|2_{+}\rangle + V_{\mathrm{DD},-}|2_{-}\rangle
\right).
\end{equation}
Under this effective basis, the molecule–atom interaction can be reduced to a single coupling channel between $|0r\rangle$ and $|2R\rangle$. The corresponding effective interaction strength is then obtained as
\begin{equation}
\frac{V}{2} =
\langle 0r|\hat{H}_{\mathrm{DD}}|2R\rangle
=
\frac{1}{2}\sqrt{V_{\mathrm{DD},+}^{2}+V_{\mathrm{DD},-}^{2}}
\approx 2\pi \times 2.02~\mathrm{MHz}.
\end{equation}
Therefore, the total interaction strength is estimated to be $V \approx 2\pi \times 4.04~\mathrm{MHz}$.

\begin{figure*}
\centering
\includegraphics[width=0.85\linewidth]{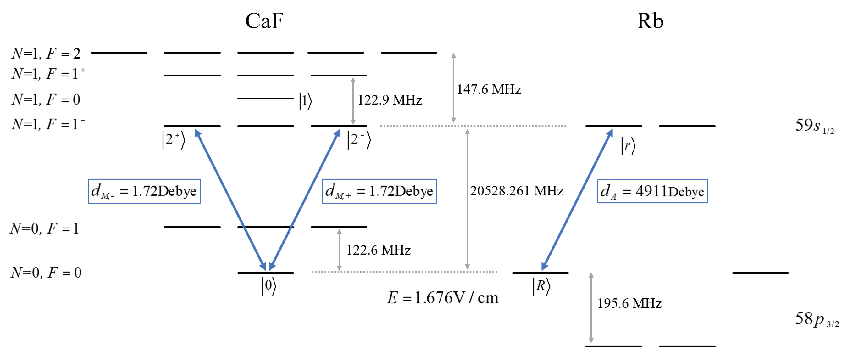}
\caption{Energy-level diagram and state selection for the $\mathrm{CaF}$ molecule and the $^{87}\mathrm{Rb}$ atom. In the presence of a static electric field $E = 1.676\,\mathrm{V/cm}$, the rotational and hyperfine levels of $\mathrm{CaF}$ are Stark shifted, and the states $|0\rangle$ and $|1\rangle$ are chosen as qubit states, coupled via intermediate states $|2^{+}\rangle$ and $|2^{-}\rangle$ with transition dipole moments $d_{M,\pm} = 1.72\,\mathrm{Debye}$. For the $^{87}\mathrm{Rb}$ atom, the Rydberg states $|R\rangle = 58p_{3/2}$ and $|r\rangle = 59s_{1/2}$ are coupled via an electric-dipole transition with dipole moment $d_A = 4911\,\mathrm{Debye}$.}
\label{F0}
\end{figure*}

\section{Effective Hamiltonian}
We derive the effective Hamiltonians $\hat H^{\textup{eff}}_1$ and $\hat H^{\textup{eff}}_2$ corresponding to two multipartite CNOT gates (CNOT) : the two-to-one CNOT gate and the one-to-two CNOT gate.

For the two-to-one CNOT gate, the system Hamiltonian is given by
\begin{align}
\hat H_{1}
=\frac{\Omega_{1}}{2} |g\rangle \langle r|+\frac{\Omega_{2}}{2} |e\rangle \langle r|+\frac{V}{2}\sum_{j=1}^{2} |0r\rangle_j \langle 2R|+ \mathrm{H.c.}
\end{align}
where $\Omega_{1}$ and $\Omega_{2}$ denote the Rabi frequencies driving the atomic transitions $|g\rangle \leftrightarrow |r\rangle$ and $|e\rangle \leftrightarrow |r\rangle$, respectively, and $V$ represents the dipole–dipole interaction strength between the molecule and the Rydberg atom.
We rewrite the Hamiltonian in the two-molecule basis as
\begin{align}
\hat H_{1}
=&\frac{\Omega_{1}}{2}\Big(
|00g\rangle \langle 00r|+|01g\rangle \langle 01r|+|02g\rangle \langle 02r|
+|10g\rangle \langle 10r| \notag\\
&\quad +|11g\rangle \langle 11r|
+|12g\rangle \langle 12r|
+|20g\rangle \langle 20r|
+|21g\rangle \langle 21r| \notag\\
&\quad +|22g\rangle \langle 22r|
\Big) \notag\\
&+\frac{\Omega_{2}}{2}\Big(
|00e\rangle \langle 00r|+|01e\rangle \langle 01r|
+|02e\rangle \langle 02r| \notag\\
&\quad +|10e\rangle \langle 10r|
+|11e\rangle \langle 11r|
+|12e\rangle \langle 12r| \notag\\
&\quad +|20e\rangle \langle 20r|
+|21e\rangle \langle 21r|
+|22e\rangle \langle 22r|
\Big) \notag\\
&+\frac{V}{2}\Big(
|00r\rangle \langle 20R|+|01r\rangle \langle 21R|
+|02r\rangle \langle 22R| \notag\\
&\quad +|00r\rangle \langle 02R|
+|10r\rangle \langle 12R|
+|20r\rangle \langle 22R|
\Big) \notag\\
&+\mathrm{H.c.}
\end{align}
Next, we consider the subspace spanned by
\[
\{|00g\rangle,|00e\rangle,|10g\rangle,|10e\rangle,
|01g\rangle,|01e\rangle,|11g\rangle,|11e\rangle\}.
\]
Within this subspace, the Hamiltonian can be decomposed as
\begin{equation}
\hat H_1=\hat H^0_1+\hat H^1_{1}+\hat H^2_{1}+\hat H^3_{1},
\end{equation}
where
\begin{equation}
\begin{aligned}
\hat H^0_1 &=\frac{\Omega_{1}}{2}|11g\rangle\langle11r|
+\frac{\Omega_{2}}{2}|11e\rangle\langle11r|
+\mathrm{H.c.},\\
\hat H^1_{1} &=\frac{\Omega_{1}}{2}|10g\rangle\langle10r|
+\frac{\Omega_{2}}{2}|10e\rangle\langle10r|
+\frac{V}{2}|10r\rangle\langle12R|
+\mathrm{H.c.},\\
\hat H^2_{1} &=\frac{\Omega_{1}}{2}|01g\rangle\langle01r|
+\frac{\Omega_{2}}{2}|01e\rangle\langle01r|
+\frac{V}{2}|01r\rangle\langle21R|
+\mathrm{H.c.},\\
\hat H^3_{1} &=\frac{\Omega_{1}}{2}|00g\rangle\langle00r|
+\frac{\Omega_{2}}{2}|00e\rangle\langle00r|
+\frac{\sqrt2 V}{2}|00r\rangle\langle\varphi|
+\mathrm{H.c.}
\end{aligned}
\end{equation}
with 
\begin{equation}
|\varphi\rangle=\frac{1}{\sqrt2}(|02R\rangle+|20R\rangle).
\end{equation}
The transitions among different subspaces are governed by the Hamiltonians 
$H_1^0$ through $H_1^3$, which include both the desired 
and undesired processes involved in the realization of the two-to-one 
CNOT gate. The Hamiltonian $\hat{H}^{1}_{1}$ can be simplified by diagonalizing the molecule–atom interaction term. In the basis 
$\{|10g\rangle, |10e\rangle, |\phi_{+}\rangle, |\phi_{-}\rangle\}$, 
the dressed states are defined as
\begin{equation}
|\phi_{\pm}\rangle=
\frac{1}{\sqrt{2}}
\left(|10r\rangle \pm |12R\rangle\right),
\end{equation}
which are the eigenstates of the molecule–atom interaction Hamiltonian 
$\frac{V}{2}(|10r\rangle\langle12R|+\mathrm{H.c.})$.
In this representation, the Hamiltonian becomes
\begin{equation}
\begin{aligned}
\hat{H}_{1} = 
&\frac{\Omega_{1}}{2\sqrt{2}}
\left( |10g\rangle\langle\phi_{+}| + |10g\rangle\langle\phi_{-}| \right)\\
&+\frac{\Omega_{2}}{2\sqrt{2}}
\left( |10e\rangle\langle\phi_{+}| + |10e\rangle\langle\phi_{-}| \right)
+ \mathrm{H.c.} \\
&+\frac{V}{2}
\left(|\phi_{+}\rangle\langle\phi_{+}|
-|\phi_{-}\rangle\langle\phi_{-}|\right).
\end{aligned}
\end{equation}
To simplify the dynamics, we perform a unitary transformation into the 
rotating frame defined by
\begin{equation}
U=e^{-i\frac{V}{2}
\left(|\phi_+\rangle\langle\phi_+|
-|\phi_-\rangle\langle\phi_-|\right)t}.
\end{equation}
Using the relation
\begin{equation}
\tilde H_1 ' = U^\dagger \hat H_{1} U - iU^\dagger \dot U,
\end{equation}
the transformed Hamiltonian becomes
\begin{equation}
\begin{aligned}
\hat H^{1}_{1} =\;&
\frac{\Omega_{1}}{2\sqrt{2}}
\left(
|10g\rangle\langle\phi_+| e^{-i\frac{V}{2}t}
+ |10g\rangle\langle\phi_-| e^{i\frac{V}{2}t}
\right) \\
&+ \frac{\Omega_{2}}{2\sqrt{2}}
\left(
|10e\rangle\langle\phi_+| e^{-i\frac{V}{2}t}
+ |10e\rangle\langle\phi_-| e^{i\frac{V}{2}t}
\right)
+ \mathrm{H.c.}
\end{aligned}
\end{equation}
Next, by leveraging the unconventional Rydberg pumping (URP) condition
\begin{equation}
V \gg \{\Omega_1,\Omega_2\},
\end{equation}
the rapidly oscillating terms can be adiabatically eliminated. 
Up to second-order perturbation, the effective Hamiltonian reduces to 
the Stark-shift form
\begin{equation}
\begin{aligned}
\hat H^{1'}_{1} =\;&
\frac{\Omega_{1}^2}{4V}
\left(|\phi_+ \rangle\langle\phi_+|
-|\phi_- \rangle\langle\phi_-|\right) \\
&+\frac{\Omega_{2}^2}{4V}
\left(|\phi_+ \rangle\langle\phi_+|
-|\phi_- \rangle\langle\phi_-|\right).
\end{aligned}
\end{equation}
which can be compensated by introducing additional ancillary levels.
In addition, the Hamiltonians $\hat{H}^{2}_{1}$ and $\hat{H}^{3}_{1}$ can 
also be reduced to Stark-shift terms under the URP condition, analogous 
to $\hat{H}^{1'}_{1}$:
\begin{equation}
\begin{aligned}
\hat H^{2'}_{1} &=
\frac{\Omega_{1}^2+\Omega_{2}^2}{4V}
\left(
|\phi_+^{\prime}\rangle\langle\phi_+^{\prime}|
-|\phi_-^{\prime}\rangle\langle\phi_-^{\prime}|
\right), \\
\hat H^{3'}_{1} &=
\frac{\Omega_{1}^2+\Omega_{2}^2}{4V}
\left(
|\psi_+\rangle\langle\psi_+|
-|\psi_-\rangle\langle\psi_-|
\right).
\end{aligned}
\end{equation}
where 
\begin{equation}
\begin{aligned}
\lvert \phi_{\pm}^{\prime}\rangle &= 
\tfrac{1}{\sqrt{2}}
\left(\lvert 01r\rangle \pm \lvert 21R\rangle\right), \\
\lvert \varphi\rangle &= 
\tfrac{1}{\sqrt{2}}
\left(\lvert 02R\rangle + \lvert 20R\rangle\right), \\
\lvert \psi_{\pm}\rangle &= 
\tfrac{1}{\sqrt{2}}
\left(\lvert 00r\rangle \pm \lvert \varphi\rangle\right).
\end{aligned}
\end{equation}
which are the eigenstates of the molecule–atom interaction Hamiltonian. 
Under the URP condition $V \gg \{\Omega_1,\Omega_2\}$, these transitions 
are far off-resonant and therefore effectively suppressed. Consequently, the effective Hamiltonian can be simplified as
\begin{align}
\hat H_1^{\mathrm{eff}}
= \hat H^0_1
=\frac{\Omega_{1}}{2}|11g\rangle\langle 11r|
+\frac{\Omega_{2}}{2}|11e\rangle\langle11r|
+ \mathrm{H.c.},
\end{align}
which describes the resonant coupling between the two-atom ground states 
and the Rydberg state $|11r\rangle$ conditioned on the molecular state 
$|1\rangle$. Therefore, the Hamiltonian $\hat H^0_1$ forms the physical basis for implementing the hybrid two-to-one CNOT gate. This conditional excitation mechanism ensures that the atomic transition 
occurs only when both molecules are in the state $|1\rangle$.

For the implementation of the one-to-two CNOT gate, two atoms are driven by classical laser fields with Rabi frequencies $\Omega_{1}$ and $\Omega_{2}$, which couple the atomic ground states $|g\rangle$ and $|e\rangle$ to the Rydberg state $|r\rangle$, respectively. Both transitions share a common detuning $\Delta$. The total Hamiltonian of the system can be expressed as
\begin{eqnarray}
\begin{aligned}
\hat H_{2} = 
&\frac{\Omega_{1}}{2}\sum_{k=1}^{2} |g\rangle_k \langle r|
+\frac{\Omega_{2}}{2}\sum_{k=1}^{2} |e\rangle_k \langle r|\\&
+\frac{V}{2}\sum_{k=1}^{2} |0r\rangle_k\langle 2R|
+\mathrm{H.c.} \\
&+U_1|rr\rangle \langle rr|
+U_2|RR\rangle \langle RR|
-\Delta\sum_{k=1}^{2}|r\rangle _k\langle r|.
\end{aligned}
\label{eq:6}
\end{eqnarray}
Here, $V$ denotes the molecule–atom dipole–dipole interaction strength, 
while $U_1$ and $U_2$ represent the Rydberg–Rydberg interaction energies 
for the states $|rr\rangle$ and $|RR\rangle$, respectively. To eliminate the rapidly oscillating phases induced by the Rydberg–Rydberg interaction terms, we perform a unitary transformation into the rotating
frame defined by
\begin{equation}
U = \exp\left[-it\left(U_{1}|rr\rangle\langle rr|
+U_{2}|RR\rangle\langle RR|\right)\right].
\end{equation}
In this frame, the transformed Hamiltonian reads
\begin{equation}
\hat H_2' = U^\dagger \hat H_{2} U - iU^\dagger \dot U .
\end{equation}
By expanding the Hamiltonian in the product basis of the two atoms and
the molecule, it can be decomposed as
\begin{equation}
\small
\setlength{\arraycolsep}{2pt}
\begin{aligned}
\hat H^0_2 =&\;
\frac{\Omega_{1}}{2}\Big(
|0gg\rangle\langle0rg|+|0gg\rangle\langle0gr|
+|0ge\rangle\langle0re|+|0eg\rangle\langle0er|\\
&\quad +|0gr\rangle\langle0rr|e^{-iU_1t}
+|0rg\rangle\langle0rr|e^{-iU_1t}
\Big)\\
&+\frac{\Omega_{2}}{2}\Big(
|0eg\rangle\langle0rg|+|0ge\rangle\langle0gr|
+|0ee\rangle\langle0re|+|0ee\rangle\langle0er|\\
&\quad +|0re\rangle\langle0rr|e^{-iU_1t}
+|0er\rangle\langle0rr|e^{-iU_1t}
\Big)\\
&+\frac{V}{2}\Big(
|0rg\rangle\langle2Rg|+|0gr\rangle\langle2gR|
+|0re\rangle\langle2Re|+|0er\rangle\langle2eR|\\
&\quad +|0rr\rangle\langle2Rr|e^{iU_1t}
+|0rr\rangle\langle2rR|e^{iU_1t}
\Big)+\mathrm{H.c.}\\
&-\Delta\Big(
|0gr\rangle\langle0gr|+|0er\rangle\langle0er|
+|0rg\rangle\langle0rg|+|0re\rangle\langle0re|
\Big)\\[6pt]
\hat H^1_2 =&\;
\frac{\Omega_{1}}{2}\Big(
|1gg\rangle\langle1rg|+|1gg\rangle\langle1gr|
+|1ge\rangle\langle1re|+|1eg\rangle\langle1er|\\
&\quad +|1gr\rangle\langle1rr|e^{-iU_1t}
+|1rg\rangle\langle1rr|e^{-iU_1t}
\Big)\\
&+\frac{\Omega_{2}}{2}\Big(
|1eg\rangle\langle1rg|+|1ge\rangle\langle1gr|
+|1ee\rangle\langle1re|+|1ee\rangle\langle1er|\\
&\quad +|1re\rangle\langle1rr|e^{-iU_1t}
+|1er\rangle\langle1rr|e^{-iU_1t}
\Big)+\mathrm{H.c.}\\
&-\Delta\Big(
|1gr\rangle\langle1gr|+|1er\rangle\langle1er|
+|1rg\rangle\langle1rg|+|1re\rangle\langle1re|
\Big)\\[6pt]
\hat H^2_2 =&\;
\frac{\Omega_{1}}{2}\Big(
|2gg\rangle\langle2rg|+|2gg\rangle\langle2gr|
+|2ge\rangle\langle2re|+|2eg\rangle\langle2er|\\
&\quad +|0gR\rangle\langle0rR|+|1gR\rangle\langle1rR|
+|0Rg\rangle\langle0Rr|+|1Rg\rangle\langle1Rr|\\
&\quad +|2gr\rangle\langle2rr|e^{-iU_1t}
+|2rg\rangle\langle2rr|e^{-iU_1t}
\Big)\\
&+\frac{\Omega_{2}}{2}\Big(
|2eg\rangle\langle2rg|+|2ge\rangle\langle2gr|
+|2ee\rangle\langle2re|+|2ee\rangle\langle2er|\\
&\quad +|0eR\rangle\langle0rR|+|1eR\rangle\langle1rR|
+|0Re\rangle\langle0Rr|+|1Re\rangle\langle1Rr|\\
&\quad +|2re\rangle\langle2rr|e^{-iU_1t}
+|2er\rangle\langle2rr|e^{-iU_1t}
\Big)\\
&+\frac{V}{2}\Big(
|0rR\rangle\langle2RR|e^{-iU_2t}
+|0Rr\rangle\langle2RR|e^{-iU_2t}
\Big)+\mathrm{H.c.}\\
&-\Delta\Big(
|2gr\rangle\langle2gr|+|2er\rangle\langle2er|
+|2Rr\rangle\langle2Rr|+|2rg\rangle\langle2rg|\\
&\quad +|2re\rangle\langle2re|
+|2rR\rangle\langle2rR|
+|0Rr\rangle\langle0Rr|
+|0rR\rangle\langle0rR|\\
&\quad +|1rR\rangle\langle1rR|
+|1Rr\rangle\langle1Rr|
\Big).
\end{aligned}
\label{2}
\end{equation}
The Hamiltonians $\hat{H}_2^0$, $\hat{H}_2^1$, and $\hat{H}_2^2$ in Eq.~(\ref{2}) correspond to the dynamical evolution of the hybrid system conditioned on the molecular states $|0\rangle$, $|1\rangle$, and $|2\rangle$, respectively. When the molecular state is initialized in $|0\rangle$, the dynamics is governed by $\hat{H}_1^0$, which couples the atomic ground states to the singly excited Rydberg states. 
When the molecular state is prepared in $|1\rangle$, the relevant
dynamical processes are described by $\hat{H}_2^1$, which accounts for
the single-excitation channels. When the molecule is in the state
$|2\rangle$, the system evolution is governed by $\hat{H}_2^2$,
including both resonant and off-resonant transitions induced by the
driving fields. Together, these Hamiltonians provide a complete description of the gate dynamics of the hybrid atom–molecule system. To analyze the gate operation, we restrict the dynamics to the computational subspace
\[
\{|0gg\rangle, |0ge\rangle, |0eg\rangle, |0ee\rangle,
|1gg\rangle, |1ge\rangle, |1eg\rangle, |1ee\rangle\},
\]
which is directly relevant for implementing the one-to-two
CNOT gate. Within this subspace, the Hamiltonian
$\hat{H}_2^2$ does not participate in the dynamics of the computational subspace and can therefore be neglected.

Under the condition $\{U_{1}, U_{2}\} \gg \{\Delta, V, \Omega_{1}, \Omega_{2}\}$, the rapidly oscillating terms associated with double Rydberg excitations can be safely neglected within the rotating-wave approximation.
When the control molecule is initially prepared in the state $|0\rangle$,
the Hamiltonian $\hat{H}_{2}^{0}$ reduces to
\begin{equation}
\resizebox{\columnwidth}{!}{$
\begin{aligned}
\hat H^0_2
=\;&\frac{\Omega_{1}}{2}\Big(
|0gg\rangle\langle0rg|+|0gg\rangle\langle0gr|
+|0ge\rangle\langle0re|+|0eg\rangle\langle0er|
\Big)\\
&+\frac{\Omega_{2}}{2}\Big(
|0eg\rangle\langle0rg|+|0ge\rangle\langle0gr|
+|0ee\rangle\langle0re|+|0ee\rangle\langle0er|
\Big)\\
&+\frac{V}{2}\Big(
|0rg\rangle\langle2Rg|+|0gr\rangle\langle2gR|
+|0re\rangle\langle2Re|+|0er\rangle\langle2eR|
\Big)
+\mathrm{H.c.}\\
&-\Delta\Big(
|0gr\rangle\langle0gr|+|0er\rangle\langle0er|
+|0rg\rangle\langle0rg|+|0re\rangle\langle0re|
\Big)
\end{aligned}
$}
\end{equation}

Furthermore, we impose the URP condition  $\Delta, V \gg \Omega_{1}, \Omega_{2}$. Under the combined effect of the strong interaction $V$ and the large detuning $\Delta$, the states 
$\{|0gr\rangle, |0rg\rangle, |0er\rangle, |0re\rangle\}$ and 
$\{|2gR\rangle, |2Rg\rangle, |2eR\rangle, |2Re\rangle\}$ 
are shifted far from resonance and form dressed states with large 
energy splittings. Because these energy shifts are much larger than the driving strengths $\Omega_{1,2}$, the transitions from the ground states to the dressed states become strongly off-resonant. As a result, the couplings between the ground states and these excited states are effectively blockaded, which suppresses unwanted excitations and confines the system dynamics within the ground-state manifold.

When the molecular qubit is prepared in the state $|1\rangle$, under the condition $\{U_{1}, U_{2}\} \gg \{\Delta, V, \Omega_{1}, \Omega_{2}\}$, this Hamiltonian can be further simplified as follows
\begin{equation}
\resizebox{\columnwidth}{!}{$
\begin{aligned}
\hat H^{1}_{2}
=\;&\frac{\Omega_{1}}{2}(
|1gg\rangle\langle1rg|
+|1gg\rangle\langle1gr|
+|1ge\rangle\langle1re|
+|1eg\rangle\langle1er|)\\
&+\frac{\Omega_{2}}{2}(
|1eg\rangle\langle1rg|
+|1ge\rangle\langle1gr|
+|1ee\rangle\langle1re|
+|1ee\rangle\langle1er|)
+\mathrm{H.c.}\\
&-\Delta(
|1gr\rangle\langle1gr|
+|1er\rangle\langle1er|
+|1rg\rangle\langle1rg|
+|1re\rangle\langle1re|)
\end{aligned}
$}
\end{equation}
During this process, the two atoms are respectively driven by laser fields with Rabi frequencies $\Omega_1$ and $\Omega_2$, which couple the atomic ground states to the Rydberg state $|r\rangle$.Under the URP condition $\Delta \gg \{\Omega_1,\Omega_2\}$, the effective 
Hamiltonian can be derived to second order in perturbation theory as
\begin{equation}
\begin{aligned}
\hat{H}^{\text{eff}}_2
=\;&\frac{\Omega_{1}\Omega_{2}}{4\Delta}
(|1gg\rangle+|1ee\rangle)(\langle1ge|+\langle1eg|)
+\mathrm{H.c.}\\
&+\frac{\Omega_{1}^2}{4\Delta}
(2|1gg\rangle\langle1gg|
+|1ge\rangle\langle1ge|
+|1eg\rangle\langle1eg|)\\
&+\frac{\Omega_{2}^2}{4\Delta}
(2|1ee\rangle\langle1ee|
+|1ge\rangle\langle1ge|
+|1eg\rangle\langle1eg|)
\end{aligned}
\end{equation}
This effective Hamiltonian governs the system dynamics within the
subspace $\{|1gg\rangle, |1ge\rangle, |1eg\rangle, |1ee\rangle\}$,
where the induced transitions occur conditionally on the molecular
qubit being prepared in the state $|1\rangle$. Such conditional
dynamics of the two atomic qubits provides the underlying physical
mechanism for implementing the one-to-two CNOT gate.

\section{Four-qubit CNOT gates}

We further extend the scheme to a four–qubit configuration in order to realize higher–order multiqubit CNOT gates, namely the three-to–one and one-to–three schemes.

\begin{figure}
\centering
\includegraphics[width=1\linewidth]{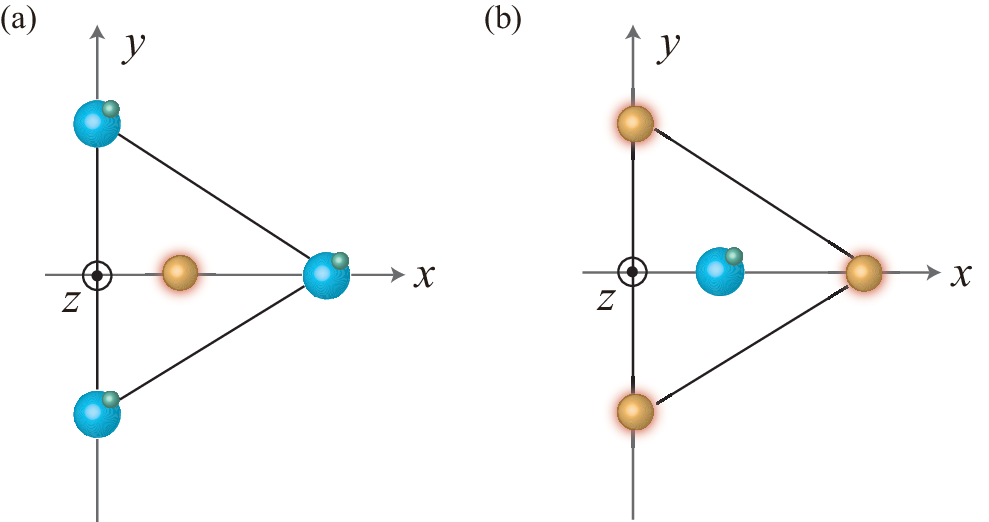}
\caption{Spatial geometric configurations of a scalable four-qubit CNOT gate.
(a) Three-to-one CNOT gate configuration: three molecular control qubits are located at the vertices of an equilateral triangle, with one atomic target qubit positioned at the geometric center.
(b) One-to-three CNOT gate configuration: three atomic target qubits are located at the vertices of an equilateral triangle, with one molecular control qubit positioned at the center.}
\label{F1}
\end{figure}

For the hybrid four-qubit CNOT gates, We consider the configuration shown in Fig. \ref{F1}: three molecules (atoms) are fixed at the vertices of an equilateral triangle, while the Rydberg atom (molecule) is located at the geometric center of the triangle. In this geometry, according to Eq. (\ref{HDD}), the molecule–atom interactions can be derived. The first molecule–atom interaction is given by
\begin{eqnarray}
	\begin{aligned}
		\frac{V_{1,+}}{2} &= \langle 0r | \hat{H}_{\text{DD}} | 2_+ R \rangle \approx -2\pi \times 0.64 \text{ MHz} ,\\
		\frac{V_{1,-}}{2} &= \langle 0r | \hat{H}_{\text{DD}} | 2_- R \rangle \approx 2\pi \times 1.91 \text{ MHz}.
	\end{aligned}
\end{eqnarray}
The second molecule–atom interaction is given by
\begin{eqnarray}
	\begin{aligned}
		\frac{V_{2,+}}{2} &= \langle 0r | \hat{H}_{\text{DD}} | 2_+ R \rangle \approx -2\pi \times 0.64 \text{ MHz} ,\\
		\frac{V_{2,-}}{2} &= \langle 0r | \hat{H}_{\text{DD}} | 2_- R \rangle \approx 2\pi \times 1.91 e^{-2i \varphi} \text{ MHz}.
	\end{aligned}
\end{eqnarray}
The third molecule–atom interaction is given by
\begin{eqnarray}
	\begin{aligned}
		\frac{V_{3,+}}{2} &= \langle 0r | \hat{H}_{\text{DD}} | 2_+ R \rangle \approx -2\pi \times 0.64 \text{ MHz} ,\\
		\frac{V_{3,-}}{2} &= \langle 0r | \hat{H}_{\text{DD}} | 2_- R \rangle \approx 2\pi \times 1.91 e^{2i \varphi} \text{ MHz},
	\end{aligned}
\end{eqnarray}
where, the azimuthal angle is $\varphi = \frac{2\pi}{3}$.

For the one-to–three CNOT gate,  as illustrated in Fig.~\ref{F1}(a), the system consists of three molecular control qubits and a single atomic target qubit. Under this configuration, the Hamiltonian of the system can be written as
\begin{equation}
\begin{aligned}
\hat{H}_{3}
=\;&\frac{\Omega_1}{2}|g\rangle\langle r|
+\frac{\Omega_2}{2}|e\rangle\langle r|\\
&+\sum_{j=1}^{3}\Big(
\frac{V_{j,+}}{2}|0r\rangle_{j}\langle2_{+}R|+\frac{V_{j,-}}{2}|0r\rangle_{j}\langle2_{-}R|
\Big)
+\mathrm{H.c.}
\end{aligned}
\label{eq:H_3to1}
\end{equation}
where the index $j$ runs over the three molecular qubits.

\begin{figure}
\centering
\includegraphics[width=1\linewidth]{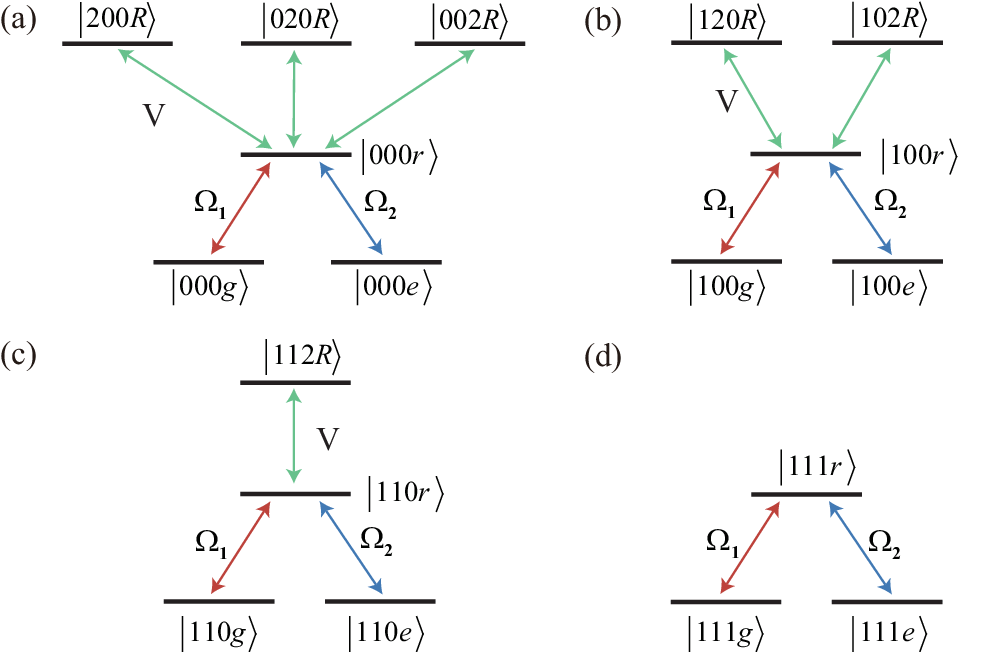}
\caption{(a) Effective transition path between $|000g\rangle$ and $|000e\rangle$.
(b) Effective transition path between $|100g\rangle$ and $|100e\rangle$.
(c) Effective transition path between $|110g\rangle$ and $|110e\rangle$.
(d) Effective transition path between $|111g\rangle$ and $|111e\rangle$. Here, the auxiliary state is defined as $|2\rangle \equiv |2_{\pm}\rangle$.}
\label{F2}
\end{figure}

To analyze the dynamical behavior of the system, we classify the sixteen computational basis states into several categories according to the molecular configurations. First, when all three molecules are prepared in the state $|0\rangle$, the relevant transition processes correspond to the basis states $|000g\rangle$ and $|000e\rangle$, as illustrated in Fig.~\ref{F2}(a). 
Second, when two of the molecules occupy the state $|0\rangle$, typical examples include the states $|100g\rangle$ and $|100e\rangle$, whose transition pathways are shown in Fig.~\ref{F2}(b). 
Similar processes also occur for the states $|010g\rangle$, $|010e\rangle$, $|001g\rangle$, and $|001e\rangle$. 
Third, when only one molecule remains in the state $|0\rangle$, the representative transitions involve the states $|110g\rangle$ and $|110e\rangle$, as depicted in Fig.~\ref{F2}(c), with analogous processes for the states $|011g\rangle$, $|011e\rangle$, $|101g\rangle$, and $|101e\rangle$.

Under the URP condition $V \gg \{\Omega_1,\Omega_2\}$, the dipole–dipole interaction term $\frac{V}{2}|0r\rangle_j\langle2R|$ leads to a significant splitting of the atomic Rydberg energy levels. 
Consequently, the corresponding excited states hybridize to form dressed states with large energy shifts. 
Because these energy shifts are much larger than the driving strengths $\Omega_{1,2}$, the transitions from the ground-state manifold to these dressed states become strongly off-resonant. 
As a result, the excitation of the atomic qubit is effectively suppressed whenever at least one molecular qubit occupies the state $|0\rangle$. Therefore, atomic excitation can only occur when all three molecular control qubits are prepared in the state $|1\rangle$. In this case, the system reduces to the subspace spanned by $|111g\rangle$ and $|111e\rangle$, and the atomic qubit can be coherently driven between the states $|g\rangle$ and $|e\rangle$ by the applied laser fields, as illustrated in Fig.~\ref{F2}(d).

Under the URP condition, the effective Hamiltonian governing the dynamics of this subspace can be derived as
\begin{equation}
\hat{H}_3^{\mathrm{eff}}
=
\frac{\Omega_1}{2}|111g\rangle\langle111r|
+
\frac{\Omega_2}{2}|111e\rangle\langle111r|
+\text{H.c.}
\label{eq:H_3to1_eff}
\end{equation}
Which clearly shows that the atomic transition occurs only when the three control qubits are simultaneously in the state $|1\rangle$. 
This conditional excitation realizes the desired three-to–one CNOT gate.

As illustrated in Fig.~\ref{Figure1}(b), the one-to–three CNOT gate consists of one molecular control qubit and three atomic target qubits. 
 In this hybrid system, besides the dipole–dipole interaction between the molecule and each atom, it is also necessary to take into account the interactions among the three Rydberg atoms. When more than one atom is excited to Rydberg state, strong van der Waals interactions arise between the atoms. These interactions play an important role in the system dynamics and must therefore be included in the theoretical description. Here, the interaction strength between the $j$th and $k$th atoms is denoted by $U_{jk}$.
The Hamiltonian of this system can therefore be written as
\begin{equation}
\begin{aligned}
\hat H_{4}
=\;&\sum_{n=1}^{3}\Big(
\frac{\Omega_{1}}{2}|g\rangle_n\langle r|
+\frac{\Omega_{2}}{2}|e\rangle_n\langle r|\\
&\quad+\frac{V_{n,+}}{2}|0r\rangle_n\langle 2_{+}R|
+\frac{V_{n,-}}{2}|0r\rangle_n\langle 2_{-}R|
+\mathrm{H.c.}
\Big)\\
&+\sum_{j\neq k}\Big(
U_{1,jk}|rr\rangle_{jk}\langle rr|
+U_{2,jk}|RR\rangle_{jk}\langle RR|
\Big)\\
&-\Delta\sum_{n=1}^{3}|r\rangle_n\langle r|.
\end{aligned}
\end{equation}
where $n$ labels the three Rydberg atoms, and $U_{1,jk}$ and $U_{2,jk}$ denote the van der Waals interaction strengths between the $j$th and $k$th atoms when they occupy the states $|rr\rangle$ and $|RR\rangle$, respectively. In the numerical simulations, we keep the same atomic states as in the previous scheme, while the corresponding interaction strengths are chosen as $U_1 = 2\pi \times 4206.4~\mathrm{MHz}$ and $U_2 = 2\pi \times 300.8~\mathrm{MHz}$.

To suppress the influence of the interatomic interactions, we employ the large-detuning regime satisfying
$\Delta \gg V \gg \{\Omega_1,\Omega_2\}$. Under this condition, the excitation of the intermediate Rydberg states becomes highly off-resonant, which effectively inhibits unwanted population transfer. 
As a consequence, the dynamics of the system is strongly dependent on the internal state of the molecular control qubit. When the molecular qubit is prepared in the state $|0\rangle$, the dipole–dipole interaction between the molecule and atoms induces significant energy shifts of the atomic Rydberg levels. 
These interaction-induced shifts lead to the formation of dressed states with large detunings, making the transitions from the atomic ground states to the excited states far from resonance. As a result, the atomic transitions are effectively blocked and the states of the three atoms remain unchanged.

In contrast, when the molecule is in the state $|1\rangle$, the dipole–dipole interaction term vanishes and the blockade mechanism is removed. In this situation, the three atoms can undergo coherent evolution driven by the laser fields. Consequently, conditional quantum transitions can occur within the atomic subsystem, allowing the implementation of the desired one-to–three CNOT operation.

\end{appendix}
\bibliography{main.bbl}

\begin{thebibliography}{88}%
\makeatletter
\providecommand \@ifxundefined [1]{%
 \@ifx{#1\undefined}
}%
\providecommand \@ifnum [1]{%
 \ifnum #1\expandafter \@firstoftwo
 \else \expandafter \@secondoftwo
 \fi
}%
\providecommand \@ifx [1]{%
 \ifx #1\expandafter \@firstoftwo
 \else \expandafter \@secondoftwo
 \fi
}%
\providecommand \natexlab [1]{#1}%
\providecommand \enquote  [1]{``#1''}%
\providecommand \bibnamefont  [1]{#1}%
\providecommand \bibfnamefont [1]{#1}%
\providecommand \citenamefont [1]{#1}%
\providecommand \href@noop [0]{\@secondoftwo}%
\providecommand \href [0]{\begingroup \@sanitize@url \@href}%
\providecommand \@href[1]{\@@startlink{#1}\@@href}%
\providecommand \@@href[1]{\endgroup#1\@@endlink}%
\providecommand \@sanitize@url [0]{\catcode `\\12\catcode `\$12\catcode `\&12\catcode `\#12\catcode `\^12\catcode `\_12\catcode `\%12\relax}%
\providecommand \@@startlink[1]{}%
\providecommand \@@endlink[0]{}%
\providecommand \url  [0]{\begingroup\@sanitize@url \@url }%
\providecommand \@url [1]{\endgroup\@href {#1}{\urlprefix }}%
\providecommand \urlprefix  [0]{URL }%
\providecommand \Eprint [0]{\href }%
\providecommand \doibase [0]{https://doi.org/}%
\providecommand \selectlanguage [0]{\@gobble}%
\providecommand \bibinfo  [0]{\@secondoftwo}%
\providecommand \bibfield  [0]{\@secondoftwo}%
\providecommand \translation [1]{[#1]}%
\providecommand \BibitemOpen [0]{}%
\providecommand \bibitemStop [0]{}%
\providecommand \bibitemNoStop [0]{.\EOS\space}%
\providecommand \EOS [0]{\spacefactor3000\relax}%
\providecommand \BibitemShut  [1]{\csname bibitem#1\endcsname}%
\let\auto@bib@innerbib\@empty
\bibitem [{\citenamefont {Nielsen}\ and\ \citenamefont {Chuang}(2010)}]{Nielsen_Chuang_2010}%
  \BibitemOpen
  \bibfield  {author} {\bibinfo {author} {\bibfnamefont {M.~A.}\ \bibnamefont {Nielsen}}\ and\ \bibinfo {author} {\bibfnamefont {I.~L.}\ \bibnamefont {Chuang}},\ }\href@noop {} {\emph {\bibinfo {title} {Quantum Computation and Quantum Information: 10th Anniversary Edition}}}\ (\bibinfo  {publisher} {Cambridge University Press},\ \bibinfo {year} {2010})\BibitemShut {NoStop}%
\bibitem [{\citenamefont {Grover}(1997)}]{PhysRevLett.79.325}%
  \BibitemOpen
  \bibfield  {author} {\bibinfo {author} {\bibfnamefont {L.~K.}\ \bibnamefont {Grover}},\ }\bibfield  {title} {\bibinfo {title} {Quantum mechanics helps in searching for a needle in a haystack},\ }\href {https://doi.org/10.1103/PhysRevLett.79.325} {\bibfield  {journal} {\bibinfo  {journal} {Phys. Rev. Lett.}\ }\textbf {\bibinfo {volume} {79}},\ \bibinfo {pages} {325} (\bibinfo {year} {1997})}\BibitemShut {NoStop}%
\bibitem [{\citenamefont {Hofmann}(2005)}]{PhysRevA.72.022329}%
  \BibitemOpen
  \bibfield  {author} {\bibinfo {author} {\bibfnamefont {H.~F.}\ \bibnamefont {Hofmann}},\ }\bibfield  {title} {\bibinfo {title} {Quantum parallelism of the controlled-not operation: An experimental criterion for the evaluation of device performance},\ }\href {https://doi.org/10.1103/PhysRevA.72.022329} {\bibfield  {journal} {\bibinfo  {journal} {Phys. Rev. A}\ }\textbf {\bibinfo {volume} {72}},\ \bibinfo {pages} {022329} (\bibinfo {year} {2005})}\BibitemShut {NoStop}%
\bibitem [{\citenamefont {Paredes}\ \emph {et~al.}(2005)\citenamefont {Paredes}, \citenamefont {Verstraete},\ and\ \citenamefont {Cirac}}]{PhysRevLett.95.140501}%
  \BibitemOpen
  \bibfield  {author} {\bibinfo {author} {\bibfnamefont {B.}~\bibnamefont {Paredes}}, \bibinfo {author} {\bibfnamefont {F.}~\bibnamefont {Verstraete}},\ and\ \bibinfo {author} {\bibfnamefont {J.~I.}\ \bibnamefont {Cirac}},\ }\bibfield  {title} {\bibinfo {title} {Exploiting quantum parallelism to simulate quantum random many-body systems},\ }\href {https://doi.org/10.1103/PhysRevLett.95.140501} {\bibfield  {journal} {\bibinfo  {journal} {Phys. Rev. Lett.}\ }\textbf {\bibinfo {volume} {95}},\ \bibinfo {pages} {140501} (\bibinfo {year} {2005})}\BibitemShut {NoStop}%
\bibitem [{\citenamefont {DiVincenzo}(1995)}]{PhysRevA.51.1015}%
  \BibitemOpen
  \bibfield  {author} {\bibinfo {author} {\bibfnamefont {D.~P.}\ \bibnamefont {DiVincenzo}},\ }\bibfield  {title} {\bibinfo {title} {Two-bit gates are universal for quantum computation},\ }\href {https://doi.org/10.1103/PhysRevA.51.1015} {\bibfield  {journal} {\bibinfo  {journal} {Phys. Rev. A}\ }\textbf {\bibinfo {volume} {51}},\ \bibinfo {pages} {1015} (\bibinfo {year} {1995})}\BibitemShut {NoStop}%
\bibitem [{\citenamefont {Barenco}\ \emph {et~al.}(1995)\citenamefont {Barenco}, \citenamefont {Bennett}, \citenamefont {Cleve}, \citenamefont {DiVincenzo}, \citenamefont {Margolus}, \citenamefont {Shor}, \citenamefont {Sleator}, \citenamefont {Smolin},\ and\ \citenamefont {Weinfurter}}]{PhysRevA.52.3457}%
  \BibitemOpen
  \bibfield  {author} {\bibinfo {author} {\bibfnamefont {A.}~\bibnamefont {Barenco}}, \bibinfo {author} {\bibfnamefont {C.~H.}\ \bibnamefont {Bennett}}, \bibinfo {author} {\bibfnamefont {R.}~\bibnamefont {Cleve}}, \bibinfo {author} {\bibfnamefont {D.~P.}\ \bibnamefont {DiVincenzo}}, \bibinfo {author} {\bibfnamefont {N.}~\bibnamefont {Margolus}}, \bibinfo {author} {\bibfnamefont {P.}~\bibnamefont {Shor}}, \bibinfo {author} {\bibfnamefont {T.}~\bibnamefont {Sleator}}, \bibinfo {author} {\bibfnamefont {J.~A.}\ \bibnamefont {Smolin}},\ and\ \bibinfo {author} {\bibfnamefont {H.}~\bibnamefont {Weinfurter}},\ }\bibfield  {title} {\bibinfo {title} {Elementary gates for quantum computation},\ }\href {https://doi.org/10.1103/PhysRevA.52.3457} {\bibfield  {journal} {\bibinfo  {journal} {Phys. Rev. A}\ }\textbf {\bibinfo {volume} {52}},\ \bibinfo {pages} {3457} (\bibinfo {year} {1995})}\BibitemShut {NoStop}%
\bibitem [{\citenamefont {Vandersypen}\ \emph {et~al.}(2001)\citenamefont {Vandersypen}, \citenamefont {Steffen}, \citenamefont {Breyta}, \citenamefont {Yannoni}, \citenamefont {Sherwood},\ and\ \citenamefont {Chuang}}]{Vandersypen2001}%
  \BibitemOpen
  \bibfield  {author} {\bibinfo {author} {\bibfnamefont {L.~M.~K.}\ \bibnamefont {Vandersypen}}, \bibinfo {author} {\bibfnamefont {M.}~\bibnamefont {Steffen}}, \bibinfo {author} {\bibfnamefont {G.}~\bibnamefont {Breyta}}, \bibinfo {author} {\bibfnamefont {C.~S.}\ \bibnamefont {Yannoni}}, \bibinfo {author} {\bibfnamefont {M.~H.}\ \bibnamefont {Sherwood}},\ and\ \bibinfo {author} {\bibfnamefont {I.~L.}\ \bibnamefont {Chuang}},\ }\bibfield  {title} {\bibinfo {title} {Experimental realization of shor's quantum factoring algorithm using nuclear magnetic resonance},\ }\href {https://doi.org/10.1038/414883a} {\bibfield  {journal} {\bibinfo  {journal} {Nature}\ }\textbf {\bibinfo {volume} {414}},\ \bibinfo {pages} {883} (\bibinfo {year} {2001})}\BibitemShut {NoStop}%
\bibitem [{\citenamefont {Joshi}\ and\ \citenamefont {Xiao}(2006)}]{PhysRevA.74.052318}%
  \BibitemOpen
  \bibfield  {author} {\bibinfo {author} {\bibfnamefont {A.}~\bibnamefont {Joshi}}\ and\ \bibinfo {author} {\bibfnamefont {M.}~\bibnamefont {Xiao}},\ }\bibfield  {title} {\bibinfo {title} {Three-qubit quantum-gate operation in a cavity qed system},\ }\href {https://doi.org/10.1103/PhysRevA.74.052318} {\bibfield  {journal} {\bibinfo  {journal} {Phys. Rev. A}\ }\textbf {\bibinfo {volume} {74}},\ \bibinfo {pages} {052318} (\bibinfo {year} {2006})}\BibitemShut {NoStop}%
\bibitem [{\citenamefont {Ionicioiu}\ \emph {et~al.}(2009)\citenamefont {Ionicioiu}, \citenamefont {Spiller},\ and\ \citenamefont {Munro}}]{PhysRevA.80.012312}%
  \BibitemOpen
  \bibfield  {author} {\bibinfo {author} {\bibfnamefont {R.}~\bibnamefont {Ionicioiu}}, \bibinfo {author} {\bibfnamefont {T.~P.}\ \bibnamefont {Spiller}},\ and\ \bibinfo {author} {\bibfnamefont {W.~J.}\ \bibnamefont {Munro}},\ }\bibfield  {title} {\bibinfo {title} {Generalized toffoli gates using qudit catalysis},\ }\href {https://doi.org/10.1103/PhysRevA.80.012312} {\bibfield  {journal} {\bibinfo  {journal} {Phys. Rev. A}\ }\textbf {\bibinfo {volume} {80}},\ \bibinfo {pages} {012312} (\bibinfo {year} {2009})}\BibitemShut {NoStop}%
\bibitem [{\citenamefont {Tang}\ \emph {et~al.}(2022)\citenamefont {Tang}, \citenamefont {Yang}, \citenamefont {Li},\ and\ \citenamefont {Shao}}]{e24101371}%
  \BibitemOpen
  \bibfield  {author} {\bibinfo {author} {\bibfnamefont {S.}~\bibnamefont {Tang}}, \bibinfo {author} {\bibfnamefont {C.}~\bibnamefont {Yang}}, \bibinfo {author} {\bibfnamefont {D.}~\bibnamefont {Li}},\ and\ \bibinfo {author} {\bibfnamefont {X.}~\bibnamefont {Shao}},\ }\bibfield  {title} {\bibinfo {title} {Implementation of quantum algorithms via fast three-rydberg-atom ccz gates},\ }\href {https://www.mdpi.com/1099-4300/24/10/1371} {\bibfield  {journal} {\bibinfo  {journal} {Entropy}\ }\textbf {\bibinfo {volume} {24}} (\bibinfo {year} {2022})}\BibitemShut {NoStop}%
\bibitem [{\citenamefont {Zhu}\ \emph {et~al.}(2025)\citenamefont {Zhu}, \citenamefont {Shang}, \citenamefont {Fan}, \citenamefont {Kou}, \citenamefont {Qu}, \citenamefont {Yan}, \citenamefont {Zhang},\ and\ \citenamefont {Wang}}]{Zhu2025}%
  \BibitemOpen
  \bibfield  {author} {\bibinfo {author} {\bibfnamefont {Y.}~\bibnamefont {Zhu}}, \bibinfo {author} {\bibfnamefont {J.}~\bibnamefont {Shang}}, \bibinfo {author} {\bibfnamefont {Y.-n.}\ \bibnamefont {Fan}}, \bibinfo {author} {\bibfnamefont {Y.}~\bibnamefont {Kou}}, \bibinfo {author} {\bibfnamefont {X.}~\bibnamefont {Qu}}, \bibinfo {author} {\bibfnamefont {X.-a.}\ \bibnamefont {Yan}}, \bibinfo {author} {\bibfnamefont {Y.}~\bibnamefont {Zhang}},\ and\ \bibinfo {author} {\bibfnamefont {F.}~\bibnamefont {Wang}},\ }\bibfield  {title} {\bibinfo {title} {Implementation of double feynman gate in high dimensional quantum systems},\ }\href {https://doi.org/10.1038/s41598-025-97002-6} {\bibfield  {journal} {\bibinfo  {journal} {Scientific Reports}\ }\textbf {\bibinfo {volume} {15}},\ \bibinfo {pages} {12184} (\bibinfo {year} {2025})}\BibitemShut {NoStop}%
\bibitem [{\citenamefont {Fujiwara}\ and\ \citenamefont {Hasegawa}(2005)}]{PhysRevA.71.012337}%
  \BibitemOpen
  \bibfield  {author} {\bibinfo {author} {\bibfnamefont {S.}~\bibnamefont {Fujiwara}}\ and\ \bibinfo {author} {\bibfnamefont {S.}~\bibnamefont {Hasegawa}},\ }\bibfield  {title} {\bibinfo {title} {General method for realizing the conditional phase-shift gate and a simulation of grover's algorithm in an ion-trap system},\ }\href {https://doi.org/10.1103/PhysRevA.71.012337} {\bibfield  {journal} {\bibinfo  {journal} {Phys. Rev. A}\ }\textbf {\bibinfo {volume} {71}},\ \bibinfo {pages} {012337} (\bibinfo {year} {2005})}\BibitemShut {NoStop}%
\bibitem [{\citenamefont {Ralph}\ \emph {et~al.}(2007)\citenamefont {Ralph}, \citenamefont {Resch},\ and\ \citenamefont {Gilchrist}}]{PhysRevA.75.022313}%
  \BibitemOpen
  \bibfield  {author} {\bibinfo {author} {\bibfnamefont {T.~C.}\ \bibnamefont {Ralph}}, \bibinfo {author} {\bibfnamefont {K.~J.}\ \bibnamefont {Resch}},\ and\ \bibinfo {author} {\bibfnamefont {A.}~\bibnamefont {Gilchrist}},\ }\bibfield  {title} {\bibinfo {title} {Efficient toffoli gates using qudits},\ }\href {https://doi.org/10.1103/PhysRevA.75.022313} {\bibfield  {journal} {\bibinfo  {journal} {Phys. Rev. A}\ }\textbf {\bibinfo {volume} {75}},\ \bibinfo {pages} {022313} (\bibinfo {year} {2007})}\BibitemShut {NoStop}%
\bibitem [{\citenamefont {Yang}\ \emph {et~al.}(2007)\citenamefont {Yang}, \citenamefont {Chen},\ and\ \citenamefont {Feng}}]{PhysRevA.76.054301}%
  \BibitemOpen
  \bibfield  {author} {\bibinfo {author} {\bibfnamefont {W.~L.}\ \bibnamefont {Yang}}, \bibinfo {author} {\bibfnamefont {C.~Y.}\ \bibnamefont {Chen}},\ and\ \bibinfo {author} {\bibfnamefont {M.}~\bibnamefont {Feng}},\ }\bibfield  {title} {\bibinfo {title} {Implementation of three-qubit grover search in cavity quantum electrodynamics},\ }\href {https://doi.org/10.1103/PhysRevA.76.054301} {\bibfield  {journal} {\bibinfo  {journal} {Phys. Rev. A}\ }\textbf {\bibinfo {volume} {76}},\ \bibinfo {pages} {054301} (\bibinfo {year} {2007})}\BibitemShut {NoStop}%
\bibitem [{\citenamefont {Yang}\ \emph {et~al.}(2010)\citenamefont {Yang}, \citenamefont {Zheng},\ and\ \citenamefont {Nori}}]{PhysRevA.82.062326}%
  \BibitemOpen
  \bibfield  {author} {\bibinfo {author} {\bibfnamefont {C.-P.}\ \bibnamefont {Yang}}, \bibinfo {author} {\bibfnamefont {S.-B.}\ \bibnamefont {Zheng}},\ and\ \bibinfo {author} {\bibfnamefont {F.}~\bibnamefont {Nori}},\ }\bibfield  {title} {\bibinfo {title} {Multiqubit tunable phase gate of one qubit simultaneously controlling $n$ qubits in a cavity},\ }\href {https://doi.org/10.1103/PhysRevA.82.062326} {\bibfield  {journal} {\bibinfo  {journal} {Phys. Rev. A}\ }\textbf {\bibinfo {volume} {82}},\ \bibinfo {pages} {062326} (\bibinfo {year} {2010})}\BibitemShut {NoStop}%
\bibitem [{\citenamefont {Maslov}(2016)}]{PhysRevA.93.022311}%
  \BibitemOpen
  \bibfield  {author} {\bibinfo {author} {\bibfnamefont {D.}~\bibnamefont {Maslov}},\ }\bibfield  {title} {\bibinfo {title} {Advantages of using relative-phase toffoli gates with an application to multiple control toffoli optimization},\ }\href {https://doi.org/10.1103/PhysRevA.93.022311} {\bibfield  {journal} {\bibinfo  {journal} {Phys. Rev. A}\ }\textbf {\bibinfo {volume} {93}},\ \bibinfo {pages} {022311} (\bibinfo {year} {2016})}\BibitemShut {NoStop}%
\bibitem [{\citenamefont {Cory}\ \emph {et~al.}(1998)\citenamefont {Cory}, \citenamefont {Price}, \citenamefont {Maas}, \citenamefont {Knill}, \citenamefont {Laflamme}, \citenamefont {Zurek}, \citenamefont {Havel},\ and\ \citenamefont {Somaroo}}]{PhysRevLett.81.2152}%
  \BibitemOpen
  \bibfield  {author} {\bibinfo {author} {\bibfnamefont {D.~G.}\ \bibnamefont {Cory}}, \bibinfo {author} {\bibfnamefont {M.~D.}\ \bibnamefont {Price}}, \bibinfo {author} {\bibfnamefont {W.}~\bibnamefont {Maas}}, \bibinfo {author} {\bibfnamefont {E.}~\bibnamefont {Knill}}, \bibinfo {author} {\bibfnamefont {R.}~\bibnamefont {Laflamme}}, \bibinfo {author} {\bibfnamefont {W.~H.}\ \bibnamefont {Zurek}}, \bibinfo {author} {\bibfnamefont {T.~F.}\ \bibnamefont {Havel}},\ and\ \bibinfo {author} {\bibfnamefont {S.~S.}\ \bibnamefont {Somaroo}},\ }\bibfield  {title} {\bibinfo {title} {Experimental quantum error correction},\ }\href {https://doi.org/10.1103/PhysRevLett.81.2152} {\bibfield  {journal} {\bibinfo  {journal} {Phys. Rev. Lett.}\ }\textbf {\bibinfo {volume} {81}},\ \bibinfo {pages} {2152} (\bibinfo {year} {1998})}\BibitemShut {NoStop}%
\bibitem [{\citenamefont {Knill}\ \emph {et~al.}(2001)\citenamefont {Knill}, \citenamefont {Laflamme}, \citenamefont {Martinez},\ and\ \citenamefont {Negrevergne}}]{PhysRevLett.86.5811}%
  \BibitemOpen
  \bibfield  {author} {\bibinfo {author} {\bibfnamefont {E.}~\bibnamefont {Knill}}, \bibinfo {author} {\bibfnamefont {R.}~\bibnamefont {Laflamme}}, \bibinfo {author} {\bibfnamefont {R.}~\bibnamefont {Martinez}},\ and\ \bibinfo {author} {\bibfnamefont {C.}~\bibnamefont {Negrevergne}},\ }\bibfield  {title} {\bibinfo {title} {Benchmarking quantum computers: The five-qubit error correcting code},\ }\href {https://doi.org/10.1103/PhysRevLett.86.5811} {\bibfield  {journal} {\bibinfo  {journal} {Phys. Rev. Lett.}\ }\textbf {\bibinfo {volume} {86}},\ \bibinfo {pages} {5811} (\bibinfo {year} {2001})}\BibitemShut {NoStop}%
\bibitem [{\citenamefont {Chiaverini}\ \emph {et~al.}(2004)\citenamefont {Chiaverini}, \citenamefont {Leibfried}, \citenamefont {Schaetz}, \citenamefont {Barrett}, \citenamefont {Blakestad}, \citenamefont {Britton}, \citenamefont {Itano}, \citenamefont {Jost}, \citenamefont {Knill}, \citenamefont {Langer}, \citenamefont {Ozeri},\ and\ \citenamefont {Wineland}}]{Chiaverini2004}%
  \BibitemOpen
  \bibfield  {author} {\bibinfo {author} {\bibfnamefont {J.}~\bibnamefont {Chiaverini}}, \bibinfo {author} {\bibfnamefont {D.}~\bibnamefont {Leibfried}}, \bibinfo {author} {\bibfnamefont {T.}~\bibnamefont {Schaetz}}, \bibinfo {author} {\bibfnamefont {M.~D.}\ \bibnamefont {Barrett}}, \bibinfo {author} {\bibfnamefont {R.~B.}\ \bibnamefont {Blakestad}}, \bibinfo {author} {\bibfnamefont {J.}~\bibnamefont {Britton}}, \bibinfo {author} {\bibfnamefont {W.~M.}\ \bibnamefont {Itano}}, \bibinfo {author} {\bibfnamefont {J.~D.}\ \bibnamefont {Jost}}, \bibinfo {author} {\bibfnamefont {E.}~\bibnamefont {Knill}}, \bibinfo {author} {\bibfnamefont {C.}~\bibnamefont {Langer}}, \bibinfo {author} {\bibfnamefont {R.}~\bibnamefont {Ozeri}},\ and\ \bibinfo {author} {\bibfnamefont {D.~J.}\ \bibnamefont {Wineland}},\ }\bibfield  {title} {\bibinfo {title} {Realization of quantum error correction},\ }\href {https://doi.org/10.1038/nature03074} {\bibfield  {journal} {\bibinfo  {journal} {Nature}\ }\textbf {\bibinfo {volume} {432}},\
  \bibinfo {pages} {602} (\bibinfo {year} {2004})}\BibitemShut {NoStop}%
\bibitem [{\citenamefont {Pittman}\ \emph {et~al.}(2005)\citenamefont {Pittman}, \citenamefont {Jacobs},\ and\ \citenamefont {Franson}}]{PhysRevA.71.052332}%
  \BibitemOpen
  \bibfield  {author} {\bibinfo {author} {\bibfnamefont {T.~B.}\ \bibnamefont {Pittman}}, \bibinfo {author} {\bibfnamefont {B.~C.}\ \bibnamefont {Jacobs}},\ and\ \bibinfo {author} {\bibfnamefont {J.~D.}\ \bibnamefont {Franson}},\ }\bibfield  {title} {\bibinfo {title} {Demonstration of quantum error correction using linear optics},\ }\href {https://doi.org/10.1103/PhysRevA.71.052332} {\bibfield  {journal} {\bibinfo  {journal} {Phys. Rev. A}\ }\textbf {\bibinfo {volume} {71}},\ \bibinfo {pages} {052332} (\bibinfo {year} {2005})}\BibitemShut {NoStop}%
\bibitem [{\citenamefont {Aoki}\ \emph {et~al.}(2009)\citenamefont {Aoki}, \citenamefont {Takahashi}, \citenamefont {Kajiya}, \citenamefont {Yoshikawa}, \citenamefont {Braunstein}, \citenamefont {van Loock},\ and\ \citenamefont {Furusawa}}]{Aoki2009}%
  \BibitemOpen
  \bibfield  {author} {\bibinfo {author} {\bibfnamefont {T.}~\bibnamefont {Aoki}}, \bibinfo {author} {\bibfnamefont {G.}~\bibnamefont {Takahashi}}, \bibinfo {author} {\bibfnamefont {T.}~\bibnamefont {Kajiya}}, \bibinfo {author} {\bibfnamefont {J.-i.}\ \bibnamefont {Yoshikawa}}, \bibinfo {author} {\bibfnamefont {S.~L.}\ \bibnamefont {Braunstein}}, \bibinfo {author} {\bibfnamefont {P.}~\bibnamefont {van Loock}},\ and\ \bibinfo {author} {\bibfnamefont {A.}~\bibnamefont {Furusawa}},\ }\bibfield  {title} {\bibinfo {title} {Quantum error correction beyond qubits},\ }\href {https://doi.org/10.1038/nphys1309} {\bibfield  {journal} {\bibinfo  {journal} {Nature Physics}\ }\textbf {\bibinfo {volume} {5}},\ \bibinfo {pages} {541} (\bibinfo {year} {2009})}\BibitemShut {NoStop}%
\bibitem [{\citenamefont {Fedorov}\ \emph {et~al.}(2012)\citenamefont {Fedorov}, \citenamefont {Steffen}, \citenamefont {Baur}, \citenamefont {da~Silva},\ and\ \citenamefont {Wallraff}}]{Fedorov2012}%
  \BibitemOpen
  \bibfield  {author} {\bibinfo {author} {\bibfnamefont {A.}~\bibnamefont {Fedorov}}, \bibinfo {author} {\bibfnamefont {L.}~\bibnamefont {Steffen}}, \bibinfo {author} {\bibfnamefont {M.}~\bibnamefont {Baur}}, \bibinfo {author} {\bibfnamefont {M.~P.}\ \bibnamefont {da~Silva}},\ and\ \bibinfo {author} {\bibfnamefont {A.}~\bibnamefont {Wallraff}},\ }\bibfield  {title} {\bibinfo {title} {Implementation of a toffoli gate with superconducting circuits},\ }\href {https://doi.org/10.1038/nature10713} {\bibfield  {journal} {\bibinfo  {journal} {Nature}\ }\textbf {\bibinfo {volume} {481}},\ \bibinfo {pages} {170} (\bibinfo {year} {2012})}\BibitemShut {NoStop}%
\bibitem [{\citenamefont {Reed}\ \emph {et~al.}(2012)\citenamefont {Reed}, \citenamefont {DiCarlo}, \citenamefont {Nigg}, \citenamefont {Sun}, \citenamefont {Frunzio}, \citenamefont {Girvin},\ and\ \citenamefont {Schoelkopf}}]{Reed2012}%
  \BibitemOpen
  \bibfield  {author} {\bibinfo {author} {\bibfnamefont {M.~D.}\ \bibnamefont {Reed}}, \bibinfo {author} {\bibfnamefont {L.}~\bibnamefont {DiCarlo}}, \bibinfo {author} {\bibfnamefont {S.~E.}\ \bibnamefont {Nigg}}, \bibinfo {author} {\bibfnamefont {L.}~\bibnamefont {Sun}}, \bibinfo {author} {\bibfnamefont {L.}~\bibnamefont {Frunzio}}, \bibinfo {author} {\bibfnamefont {S.~M.}\ \bibnamefont {Girvin}},\ and\ \bibinfo {author} {\bibfnamefont {R.~J.}\ \bibnamefont {Schoelkopf}},\ }\bibfield  {title} {\bibinfo {title} {Realization of three-qubit quantum error correction with superconducting circuits},\ }\href {https://doi.org/10.1038/nature10786} {\bibfield  {journal} {\bibinfo  {journal} {Nature}\ }\textbf {\bibinfo {volume} {482}},\ \bibinfo {pages} {382} (\bibinfo {year} {2012})}\BibitemShut {NoStop}%
\bibitem [{\citenamefont {Stojanovi\ifmmode~\acute{c}\else \'{c}\fi{}}\ \emph {et~al.}(2012)\citenamefont {Stojanovi\ifmmode~\acute{c}\else \'{c}\fi{}}, \citenamefont {Fedorov}, \citenamefont {Wallraff},\ and\ \citenamefont {Bruder}}]{PhysRevB.85.054504}%
  \BibitemOpen
  \bibfield  {author} {\bibinfo {author} {\bibfnamefont {V.~M.}\ \bibnamefont {Stojanovi\ifmmode~\acute{c}\else \'{c}\fi{}}}, \bibinfo {author} {\bibfnamefont {A.}~\bibnamefont {Fedorov}}, \bibinfo {author} {\bibfnamefont {A.}~\bibnamefont {Wallraff}},\ and\ \bibinfo {author} {\bibfnamefont {C.}~\bibnamefont {Bruder}},\ }\bibfield  {title} {\bibinfo {title} {Quantum-control approach to realizing a toffoli gate in circuit qed},\ }\href {https://doi.org/10.1103/PhysRevB.85.054504} {\bibfield  {journal} {\bibinfo  {journal} {Phys. Rev. B}\ }\textbf {\bibinfo {volume} {85}},\ \bibinfo {pages} {054504} (\bibinfo {year} {2012})}\BibitemShut {NoStop}%
\bibitem [{\citenamefont {Zahedinejad}\ \emph {et~al.}(2015)\citenamefont {Zahedinejad}, \citenamefont {Ghosh},\ and\ \citenamefont {Sanders}}]{PhysRevLett.114.200502}%
  \BibitemOpen
  \bibfield  {author} {\bibinfo {author} {\bibfnamefont {E.}~\bibnamefont {Zahedinejad}}, \bibinfo {author} {\bibfnamefont {J.}~\bibnamefont {Ghosh}},\ and\ \bibinfo {author} {\bibfnamefont {B.~C.}\ \bibnamefont {Sanders}},\ }\bibfield  {title} {\bibinfo {title} {High-fidelity single-shot toffoli gate via quantum control},\ }\href {https://doi.org/10.1103/PhysRevLett.114.200502} {\bibfield  {journal} {\bibinfo  {journal} {Phys. Rev. Lett.}\ }\textbf {\bibinfo {volume} {114}},\ \bibinfo {pages} {200502} (\bibinfo {year} {2015})}\BibitemShut {NoStop}%
\bibitem [{\citenamefont {Zahedinejad}\ \emph {et~al.}(2016)\citenamefont {Zahedinejad}, \citenamefont {Ghosh},\ and\ \citenamefont {Sanders}}]{PhysRevApplied.6.054005}%
  \BibitemOpen
  \bibfield  {author} {\bibinfo {author} {\bibfnamefont {E.}~\bibnamefont {Zahedinejad}}, \bibinfo {author} {\bibfnamefont {J.}~\bibnamefont {Ghosh}},\ and\ \bibinfo {author} {\bibfnamefont {B.~C.}\ \bibnamefont {Sanders}},\ }\bibfield  {title} {\bibinfo {title} {Designing high-fidelity single-shot three-qubit gates: A machine-learning approach},\ }\href {https://doi.org/10.1103/PhysRevApplied.6.054005} {\bibfield  {journal} {\bibinfo  {journal} {Phys. Rev. Appl.}\ }\textbf {\bibinfo {volume} {6}},\ \bibinfo {pages} {054005} (\bibinfo {year} {2016})}\BibitemShut {NoStop}%
\bibitem [{\citenamefont {Roy}\ \emph {et~al.}(2020)\citenamefont {Roy}, \citenamefont {Hazra}, \citenamefont {Kundu}, \citenamefont {Chand}, \citenamefont {Patankar},\ and\ \citenamefont {Vijay}}]{PhysRevApplied.14.014072}%
  \BibitemOpen
  \bibfield  {author} {\bibinfo {author} {\bibfnamefont {T.}~\bibnamefont {Roy}}, \bibinfo {author} {\bibfnamefont {S.}~\bibnamefont {Hazra}}, \bibinfo {author} {\bibfnamefont {S.}~\bibnamefont {Kundu}}, \bibinfo {author} {\bibfnamefont {M.}~\bibnamefont {Chand}}, \bibinfo {author} {\bibfnamefont {M.~P.}\ \bibnamefont {Patankar}},\ and\ \bibinfo {author} {\bibfnamefont {R.}~\bibnamefont {Vijay}},\ }\bibfield  {title} {\bibinfo {title} {Programmable superconducting processor with native three-qubit gates},\ }\href {https://doi.org/10.1103/PhysRevApplied.14.014072} {\bibfield  {journal} {\bibinfo  {journal} {Phys. Rev. Appl.}\ }\textbf {\bibinfo {volume} {14}},\ \bibinfo {pages} {014072} (\bibinfo {year} {2020})}\BibitemShut {NoStop}%
\bibitem [{\citenamefont {Li}\ \emph {et~al.}(2024)\citenamefont {Li}, \citenamefont {Tao}, \citenamefont {Yi}, \citenamefont {Luo}, \citenamefont {Zhang}, \citenamefont {Zhou}, \citenamefont {Liu}, \citenamefont {Yan}, \citenamefont {Chen},\ and\ \citenamefont {Yu}}]{Xiao-LeLi:51205}%
  \BibitemOpen
  \bibfield  {author} {\bibinfo {author} {\bibfnamefont {X.-L.}\ \bibnamefont {Li}}, \bibinfo {author} {\bibfnamefont {Z.}~\bibnamefont {Tao}}, \bibinfo {author} {\bibfnamefont {K.}~\bibnamefont {Yi}}, \bibinfo {author} {\bibfnamefont {K.}~\bibnamefont {Luo}}, \bibinfo {author} {\bibfnamefont {L.}~\bibnamefont {Zhang}}, \bibinfo {author} {\bibfnamefont {Y.}~\bibnamefont {Zhou}}, \bibinfo {author} {\bibfnamefont {S.}~\bibnamefont {Liu}}, \bibinfo {author} {\bibfnamefont {T.}~\bibnamefont {Yan}}, \bibinfo {author} {\bibfnamefont {Y.}~\bibnamefont {Chen}},\ and\ \bibinfo {author} {\bibfnamefont {D.}~\bibnamefont {Yu}},\ }\bibfield  {title} {\bibinfo {title} {Hardware-efficient and fast three-qubit gate in superconducting quantum circuits},\ }\href {https://doi.org/10.1007/s11467-024-1405-8} {\bibfield  {journal} {\bibinfo  {journal} {Frontiers of Physics}\ }\textbf {\bibinfo {volume} {19}},\ \bibinfo {eid} {51205} (\bibinfo {year} {2024})}\BibitemShut {NoStop}%
\bibitem [{\citenamefont {Wang}\ \emph {et~al.}(2001)\citenamefont {Wang}, \citenamefont {S\o{}rensen},\ and\ \citenamefont {M\o{}lmer}}]{PhysRevLett.86.3907}%
  \BibitemOpen
  \bibfield  {author} {\bibinfo {author} {\bibfnamefont {X.}~\bibnamefont {Wang}}, \bibinfo {author} {\bibfnamefont {A.}~\bibnamefont {S\o{}rensen}},\ and\ \bibinfo {author} {\bibfnamefont {K.}~\bibnamefont {M\o{}lmer}},\ }\bibfield  {title} {\bibinfo {title} {Multibit gates for quantum computing},\ }\href {https://doi.org/10.1103/PhysRevLett.86.3907} {\bibfield  {journal} {\bibinfo  {journal} {Phys. Rev. Lett.}\ }\textbf {\bibinfo {volume} {86}},\ \bibinfo {pages} {3907} (\bibinfo {year} {2001})}\BibitemShut {NoStop}%
\bibitem [{\citenamefont {Monz}\ \emph {et~al.}(2009)\citenamefont {Monz}, \citenamefont {Kim}, \citenamefont {H\"ansel}, \citenamefont {Riebe}, \citenamefont {Villar}, \citenamefont {Schindler}, \citenamefont {Chwalla}, \citenamefont {Hennrich},\ and\ \citenamefont {Blatt}}]{PhysRevLett.102.040501}%
  \BibitemOpen
  \bibfield  {author} {\bibinfo {author} {\bibfnamefont {T.}~\bibnamefont {Monz}}, \bibinfo {author} {\bibfnamefont {K.}~\bibnamefont {Kim}}, \bibinfo {author} {\bibfnamefont {W.}~\bibnamefont {H\"ansel}}, \bibinfo {author} {\bibfnamefont {M.}~\bibnamefont {Riebe}}, \bibinfo {author} {\bibfnamefont {A.~S.}\ \bibnamefont {Villar}}, \bibinfo {author} {\bibfnamefont {P.}~\bibnamefont {Schindler}}, \bibinfo {author} {\bibfnamefont {M.}~\bibnamefont {Chwalla}}, \bibinfo {author} {\bibfnamefont {M.}~\bibnamefont {Hennrich}},\ and\ \bibinfo {author} {\bibfnamefont {R.}~\bibnamefont {Blatt}},\ }\bibfield  {title} {\bibinfo {title} {Realization of the quantum toffoli gate with trapped ions},\ }\href {https://doi.org/10.1103/PhysRevLett.102.040501} {\bibfield  {journal} {\bibinfo  {journal} {Phys. Rev. Lett.}\ }\textbf {\bibinfo {volume} {102}},\ \bibinfo {pages} {040501} (\bibinfo {year} {2009})}\BibitemShut {NoStop}%
\bibitem [{\citenamefont {Borrelli}\ \emph {et~al.}(2011)\citenamefont {Borrelli}, \citenamefont {Mazzola}, \citenamefont {Paternostro},\ and\ \citenamefont {Maniscalco}}]{PhysRevA.84.012314}%
  \BibitemOpen
  \bibfield  {author} {\bibinfo {author} {\bibfnamefont {M.}~\bibnamefont {Borrelli}}, \bibinfo {author} {\bibfnamefont {L.}~\bibnamefont {Mazzola}}, \bibinfo {author} {\bibfnamefont {M.}~\bibnamefont {Paternostro}},\ and\ \bibinfo {author} {\bibfnamefont {S.}~\bibnamefont {Maniscalco}},\ }\bibfield  {title} {\bibinfo {title} {Simple trapped-ion architecture for high-fidelity toffoli gates},\ }\href {https://doi.org/10.1103/PhysRevA.84.012314} {\bibfield  {journal} {\bibinfo  {journal} {Phys. Rev. A}\ }\textbf {\bibinfo {volume} {84}},\ \bibinfo {pages} {012314} (\bibinfo {year} {2011})}\BibitemShut {NoStop}%
\bibitem [{\citenamefont {Figgatt}\ \emph {et~al.}(2017)\citenamefont {Figgatt}, \citenamefont {Maslov}, \citenamefont {Landsman}, \citenamefont {Linke}, \citenamefont {Debnath},\ and\ \citenamefont {Monroe}}]{Figgatt2017}%
  \BibitemOpen
  \bibfield  {author} {\bibinfo {author} {\bibfnamefont {C.}~\bibnamefont {Figgatt}}, \bibinfo {author} {\bibfnamefont {D.}~\bibnamefont {Maslov}}, \bibinfo {author} {\bibfnamefont {K.~A.}\ \bibnamefont {Landsman}}, \bibinfo {author} {\bibfnamefont {N.~M.}\ \bibnamefont {Linke}}, \bibinfo {author} {\bibfnamefont {S.}~\bibnamefont {Debnath}},\ and\ \bibinfo {author} {\bibfnamefont {C.}~\bibnamefont {Monroe}},\ }\bibfield  {title} {\bibinfo {title} {Complete 3-qubit grover search on a programmable quantum computer},\ }\href {https://doi.org/10.1038/s41467-017-01904-7} {\bibfield  {journal} {\bibinfo  {journal} {Nature Communications}\ }\textbf {\bibinfo {volume} {8}},\ \bibinfo {pages} {1918} (\bibinfo {year} {2017})}\BibitemShut {NoStop}%
\bibitem [{\citenamefont {Milburn}(1989)}]{PhysRevLett.62.2124}%
  \BibitemOpen
  \bibfield  {author} {\bibinfo {author} {\bibfnamefont {G.~J.}\ \bibnamefont {Milburn}},\ }\bibfield  {title} {\bibinfo {title} {Quantum optical fredkin gate},\ }\href {https://doi.org/10.1103/PhysRevLett.62.2124} {\bibfield  {journal} {\bibinfo  {journal} {Phys. Rev. Lett.}\ }\textbf {\bibinfo {volume} {62}},\ \bibinfo {pages} {2124} (\bibinfo {year} {1989})}\BibitemShut {NoStop}%
\bibitem [{\citenamefont {Fiur\'a\ifmmode~\check{s}\else \v{s}\fi{}ek}(2008)}]{PhysRevA.78.032317}%
  \BibitemOpen
  \bibfield  {author} {\bibinfo {author} {\bibfnamefont {J.}~\bibnamefont {Fiur\'a\ifmmode~\check{s}\else \v{s}\fi{}ek}},\ }\bibfield  {title} {\bibinfo {title} {Linear optical fredkin gate based on partial-swap gate},\ }\href {https://doi.org/10.1103/PhysRevA.78.032317} {\bibfield  {journal} {\bibinfo  {journal} {Phys. Rev. A}\ }\textbf {\bibinfo {volume} {78}},\ \bibinfo {pages} {032317} (\bibinfo {year} {2008})}\BibitemShut {NoStop}%
\bibitem [{\citenamefont {Lanyon}\ \emph {et~al.}(2009)\citenamefont {Lanyon}, \citenamefont {Barbieri}, \citenamefont {Almeida}, \citenamefont {Jennewein}, \citenamefont {Ralph}, \citenamefont {Resch}, \citenamefont {Pryde}, \citenamefont {O'Brien}, \citenamefont {Gilchrist},\ and\ \citenamefont {White}}]{Lanyon2009}%
  \BibitemOpen
  \bibfield  {author} {\bibinfo {author} {\bibfnamefont {B.~P.}\ \bibnamefont {Lanyon}}, \bibinfo {author} {\bibfnamefont {M.}~\bibnamefont {Barbieri}}, \bibinfo {author} {\bibfnamefont {M.~P.}\ \bibnamefont {Almeida}}, \bibinfo {author} {\bibfnamefont {T.}~\bibnamefont {Jennewein}}, \bibinfo {author} {\bibfnamefont {T.~C.}\ \bibnamefont {Ralph}}, \bibinfo {author} {\bibfnamefont {K.~J.}\ \bibnamefont {Resch}}, \bibinfo {author} {\bibfnamefont {G.~J.}\ \bibnamefont {Pryde}}, \bibinfo {author} {\bibfnamefont {J.~L.}\ \bibnamefont {O'Brien}}, \bibinfo {author} {\bibfnamefont {A.}~\bibnamefont {Gilchrist}},\ and\ \bibinfo {author} {\bibfnamefont {A.~G.}\ \bibnamefont {White}},\ }\bibfield  {title} {\bibinfo {title} {Simplifying quantum logic using higher-dimensional hilbert spaces},\ }\href {https://doi.org/10.1038/nphys1150} {\bibfield  {journal} {\bibinfo  {journal} {Nature Physics}\ }\textbf {\bibinfo {volume} {5}},\ \bibinfo {pages} {134} (\bibinfo {year} {2009})}\BibitemShut {NoStop}%
\bibitem [{\citenamefont {Lin}\ and\ \citenamefont {He}(2009)}]{PhysRevA.80.042310}%
  \BibitemOpen
  \bibfield  {author} {\bibinfo {author} {\bibfnamefont {Q.}~\bibnamefont {Lin}}\ and\ \bibinfo {author} {\bibfnamefont {B.}~\bibnamefont {He}},\ }\bibfield  {title} {\bibinfo {title} {Single-photon logic gates using minimal resources},\ }\href {https://doi.org/10.1103/PhysRevA.80.042310} {\bibfield  {journal} {\bibinfo  {journal} {Phys. Rev. A}\ }\textbf {\bibinfo {volume} {80}},\ \bibinfo {pages} {042310} (\bibinfo {year} {2009})}\BibitemShut {NoStop}%
\bibitem [{\citenamefont {Mi\ifmmode~\check{c}\else \v{c}\fi{}uda}\ \emph {et~al.}(2013)\citenamefont {Mi\ifmmode~\check{c}\else \v{c}\fi{}uda}, \citenamefont {Sedl\'ak}, \citenamefont {Straka}, \citenamefont {Mikov\'a}, \citenamefont {Du\ifmmode~\check{s}\else \v{s}\fi{}ek}, \citenamefont {Je\ifmmode~\check{z}\else \v{z}\fi{}ek},\ and\ \citenamefont {Fiur\'a\ifmmode~\check{s}\else \v{s}\fi{}ek}}]{PhysRevLett.111.160407}%
  \BibitemOpen
  \bibfield  {author} {\bibinfo {author} {\bibfnamefont {M.}~\bibnamefont {Mi\ifmmode~\check{c}\else \v{c}\fi{}uda}}, \bibinfo {author} {\bibfnamefont {M.}~\bibnamefont {Sedl\'ak}}, \bibinfo {author} {\bibfnamefont {I.}~\bibnamefont {Straka}}, \bibinfo {author} {\bibfnamefont {M.}~\bibnamefont {Mikov\'a}}, \bibinfo {author} {\bibfnamefont {M.}~\bibnamefont {Du\ifmmode~\check{s}\else \v{s}\fi{}ek}}, \bibinfo {author} {\bibfnamefont {M.}~\bibnamefont {Je\ifmmode~\check{z}\else \v{z}\fi{}ek}},\ and\ \bibinfo {author} {\bibfnamefont {J.}~\bibnamefont {Fiur\'a\ifmmode~\check{s}\else \v{s}\fi{}ek}},\ }\bibfield  {title} {\bibinfo {title} {Efficient experimental estimation of fidelity of linear optical quantum toffoli gate},\ }\href {https://doi.org/10.1103/PhysRevLett.111.160407} {\bibfield  {journal} {\bibinfo  {journal} {Phys. Rev. Lett.}\ }\textbf {\bibinfo {volume} {111}},\ \bibinfo {pages} {160407} (\bibinfo {year} {2013})}\BibitemShut {NoStop}%
\bibitem [{\citenamefont {Liu}\ and\ \citenamefont {Wei}(2020)}]{Liu_2020}%
  \BibitemOpen
  \bibfield  {author} {\bibinfo {author} {\bibfnamefont {W.-Q.}\ \bibnamefont {Liu}}\ and\ \bibinfo {author} {\bibfnamefont {H.-R.}\ \bibnamefont {Wei}},\ }\bibfield  {title} {\bibinfo {title} {Optimal synthesis of the fredkin gate in a multilevel system},\ }\href {https://doi.org/10.1088/1367-2630/ab8e13} {\bibfield  {journal} {\bibinfo  {journal} {New Journal of Physics}\ }\textbf {\bibinfo {volume} {22}},\ \bibinfo {pages} {063026} (\bibinfo {year} {2020})}\BibitemShut {NoStop}%
\bibitem [{\citenamefont {Li}\ \emph {et~al.}(2022)\citenamefont {Li}, \citenamefont {Wan}, \citenamefont {Zhang}, \citenamefont {Zhu}, \citenamefont {Shi}, \citenamefont {Chin}, \citenamefont {Zhou}, \citenamefont {Kwek},\ and\ \citenamefont {Liu}}]{Li2022}%
  \BibitemOpen
  \bibfield  {author} {\bibinfo {author} {\bibfnamefont {Y.}~\bibnamefont {Li}}, \bibinfo {author} {\bibfnamefont {L.}~\bibnamefont {Wan}}, \bibinfo {author} {\bibfnamefont {H.}~\bibnamefont {Zhang}}, \bibinfo {author} {\bibfnamefont {H.}~\bibnamefont {Zhu}}, \bibinfo {author} {\bibfnamefont {Y.}~\bibnamefont {Shi}}, \bibinfo {author} {\bibfnamefont {L.~K.}\ \bibnamefont {Chin}}, \bibinfo {author} {\bibfnamefont {X.}~\bibnamefont {Zhou}}, \bibinfo {author} {\bibfnamefont {L.~C.}\ \bibnamefont {Kwek}},\ and\ \bibinfo {author} {\bibfnamefont {A.~Q.}\ \bibnamefont {Liu}},\ }\bibfield  {title} {\bibinfo {title} {Quantum fredkin and toffoli gates on a versatile programmable silicon photonic chip},\ }\href {https://doi.org/10.1038/s41534-022-00627-y} {\bibfield  {journal} {\bibinfo  {journal} {npj Quantum Information}\ }\textbf {\bibinfo {volume} {8}},\ \bibinfo {pages} {112} (\bibinfo {year} {2022})}\BibitemShut {NoStop}%
\bibitem [{\citenamefont {Baldazzi}\ and\ \citenamefont {Pavesi}(2025)}]{https://doi.org/10.1002/qute.202400418}%
  \BibitemOpen
  \bibfield  {author} {\bibinfo {author} {\bibfnamefont {A.}~\bibnamefont {Baldazzi}}\ and\ \bibinfo {author} {\bibfnamefont {L.}~\bibnamefont {Pavesi}},\ }\bibfield  {title} {\bibinfo {title} {Universal multiport interferometers for post-selected multi-photon gates},\ }\href {https://doi.org/https://doi.org/10.1002/qute.202400418} {\bibfield  {journal} {\bibinfo  {journal} {Advanced Quantum Technologies}\ }\textbf {\bibinfo {volume} {8}},\ \bibinfo {pages} {2400418} (\bibinfo {year} {2025})}\BibitemShut {NoStop}%
\bibitem [{\citenamefont {Toozandehjani}\ \emph {et~al.}(2025)\citenamefont {Toozandehjani}, \citenamefont {Khosroabadi},\ and\ \citenamefont {Houshmand}}]{Toozandehjani2025}%
  \BibitemOpen
  \bibfield  {author} {\bibinfo {author} {\bibfnamefont {H.}~\bibnamefont {Toozandehjani}}, \bibinfo {author} {\bibfnamefont {S.}~\bibnamefont {Khosroabadi}},\ and\ \bibinfo {author} {\bibfnamefont {M.}~\bibnamefont {Houshmand}},\ }\bibfield  {title} {\bibinfo {title} {A novel design of high contrast ratio quantum c2not (toffoli) gate based on photonic crystals},\ }\href {https://doi.org/10.1007/s11128-025-04673-1} {\bibfield  {journal} {\bibinfo  {journal} {Quantum Information Processing}\ }\textbf {\bibinfo {volume} {24}},\ \bibinfo {pages} {57} (\bibinfo {year} {2025})}\BibitemShut {NoStop}%
\bibitem [{\citenamefont {Gullans}\ and\ \citenamefont {Petta}(2019)}]{PhysRevB.100.085419}%
  \BibitemOpen
  \bibfield  {author} {\bibinfo {author} {\bibfnamefont {M.~J.}\ \bibnamefont {Gullans}}\ and\ \bibinfo {author} {\bibfnamefont {J.~R.}\ \bibnamefont {Petta}},\ }\bibfield  {title} {\bibinfo {title} {Protocol for a resonantly driven three-qubit toffoli gate with silicon spin qubits},\ }\href {https://doi.org/10.1103/PhysRevB.100.085419} {\bibfield  {journal} {\bibinfo  {journal} {Phys. Rev. B}\ }\textbf {\bibinfo {volume} {100}},\ \bibinfo {pages} {085419} (\bibinfo {year} {2019})}\BibitemShut {NoStop}%
\bibitem [{\citenamefont {Takeda}\ \emph {et~al.}(2022)\citenamefont {Takeda}, \citenamefont {Noiri}, \citenamefont {Nakajima}, \citenamefont {Kobayashi},\ and\ \citenamefont {Tarucha}}]{Takeda2022}%
  \BibitemOpen
  \bibfield  {author} {\bibinfo {author} {\bibfnamefont {K.}~\bibnamefont {Takeda}}, \bibinfo {author} {\bibfnamefont {A.}~\bibnamefont {Noiri}}, \bibinfo {author} {\bibfnamefont {T.}~\bibnamefont {Nakajima}}, \bibinfo {author} {\bibfnamefont {T.}~\bibnamefont {Kobayashi}},\ and\ \bibinfo {author} {\bibfnamefont {S.}~\bibnamefont {Tarucha}},\ }\bibfield  {title} {\bibinfo {title} {Quantum error correction with silicon spin qubits},\ }\href {https://doi.org/10.1038/s41586-022-04986-6} {\bibfield  {journal} {\bibinfo  {journal} {Nature}\ }\textbf {\bibinfo {volume} {608}},\ \bibinfo {pages} {682} (\bibinfo {year} {2022})}\BibitemShut {NoStop}%
\bibitem [{\citenamefont {Zhou}\ \emph {et~al.}(2025)\citenamefont {Zhou}, \citenamefont {He}, \citenamefont {Pang}, \citenamefont {Lyu}, \citenamefont {Zhang},\ and\ \citenamefont {Chen}}]{PhysRevA.111.042616}%
  \BibitemOpen
  \bibfield  {author} {\bibinfo {author} {\bibfnamefont {Y.}~\bibnamefont {Zhou}}, \bibinfo {author} {\bibfnamefont {H.}~\bibnamefont {He}}, \bibinfo {author} {\bibfnamefont {F.}~\bibnamefont {Pang}}, \bibinfo {author} {\bibfnamefont {H.}~\bibnamefont {Lyu}}, \bibinfo {author} {\bibfnamefont {Y.}~\bibnamefont {Zhang}},\ and\ \bibinfo {author} {\bibfnamefont {X.}~\bibnamefont {Chen}},\ }\bibfield  {title} {\bibinfo {title} {Variational quantum compiling for three-qubit-gate design in quantum dots},\ }\href {https://doi.org/10.1103/PhysRevA.111.042616} {\bibfield  {journal} {\bibinfo  {journal} {Phys. Rev. A}\ }\textbf {\bibinfo {volume} {111}},\ \bibinfo {pages} {042616} (\bibinfo {year} {2025})}\BibitemShut {NoStop}%
\bibitem [{\citenamefont {Duan}\ \emph {et~al.}(2005)\citenamefont {Duan}, \citenamefont {Wang},\ and\ \citenamefont {Kimble}}]{PhysRevA.72.032333}%
  \BibitemOpen
  \bibfield  {author} {\bibinfo {author} {\bibfnamefont {L.-M.}\ \bibnamefont {Duan}}, \bibinfo {author} {\bibfnamefont {B.}~\bibnamefont {Wang}},\ and\ \bibinfo {author} {\bibfnamefont {H.~J.}\ \bibnamefont {Kimble}},\ }\bibfield  {title} {\bibinfo {title} {Robust quantum gates on neutral atoms with cavity-assisted photon scattering},\ }\href {https://doi.org/10.1103/PhysRevA.72.032333} {\bibfield  {journal} {\bibinfo  {journal} {Phys. Rev. A}\ }\textbf {\bibinfo {volume} {72}},\ \bibinfo {pages} {032333} (\bibinfo {year} {2005})}\BibitemShut {NoStop}%
\bibitem [{\citenamefont {Isenhower}\ \emph {et~al.}(2011)\citenamefont {Isenhower}, \citenamefont {Saffman},\ and\ \citenamefont {M{\o}lmer}}]{Isenhower2011}%
  \BibitemOpen
  \bibfield  {author} {\bibinfo {author} {\bibfnamefont {L.}~\bibnamefont {Isenhower}}, \bibinfo {author} {\bibfnamefont {M.}~\bibnamefont {Saffman}},\ and\ \bibinfo {author} {\bibfnamefont {K.}~\bibnamefont {M{\o}lmer}},\ }\bibfield  {title} {\bibinfo {title} {Multibit cknot quantum gates via rydberg blockade},\ }\href {https://doi.org/10.1007/s11128-011-0292-4} {\bibfield  {journal} {\bibinfo  {journal} {Quantum Information Processing}\ }\textbf {\bibinfo {volume} {10}},\ \bibinfo {pages} {755} (\bibinfo {year} {2011})}\BibitemShut {NoStop}%
\bibitem [{\citenamefont {Shi}(2018)}]{PhysRevApplied.9.051001}%
  \BibitemOpen
  \bibfield  {author} {\bibinfo {author} {\bibfnamefont {X.-F.}\ \bibnamefont {Shi}},\ }\bibfield  {title} {\bibinfo {title} {Deutsch, toffoli, and cnot gates via rydberg blockade of neutral atoms},\ }\href {https://doi.org/10.1103/PhysRevApplied.9.051001} {\bibfield  {journal} {\bibinfo  {journal} {Phys. Rev. Appl.}\ }\textbf {\bibinfo {volume} {9}},\ \bibinfo {pages} {051001} (\bibinfo {year} {2018})}\BibitemShut {NoStop}%
\bibitem [{\citenamefont {Li}\ and\ \citenamefont {Shao}(2018)}]{PhysRevA.98.062338}%
  \BibitemOpen
  \bibfield  {author} {\bibinfo {author} {\bibfnamefont {D.~X.}\ \bibnamefont {Li}}\ and\ \bibinfo {author} {\bibfnamefont {X.~Q.}\ \bibnamefont {Shao}},\ }\bibfield  {title} {\bibinfo {title} {Unconventional rydberg pumping and applications in quantum information processing},\ }\href {https://doi.org/10.1103/PhysRevA.98.062338} {\bibfield  {journal} {\bibinfo  {journal} {Phys. Rev. A}\ }\textbf {\bibinfo {volume} {98}},\ \bibinfo {pages} {062338} (\bibinfo {year} {2018})}\BibitemShut {NoStop}%
\bibitem [{\citenamefont {Beterov}\ \emph {et~al.}(2018)\citenamefont {Beterov}, \citenamefont {Ashkarin}, \citenamefont {Yakshina}, \citenamefont {Tretyakov}, \citenamefont {Entin}, \citenamefont {Ryabtsev}, \citenamefont {Cheinet}, \citenamefont {Pillet},\ and\ \citenamefont {Saffman}}]{PhysRevA.98.042704}%
  \BibitemOpen
  \bibfield  {author} {\bibinfo {author} {\bibfnamefont {I.~I.}\ \bibnamefont {Beterov}}, \bibinfo {author} {\bibfnamefont {I.~N.}\ \bibnamefont {Ashkarin}}, \bibinfo {author} {\bibfnamefont {E.~A.}\ \bibnamefont {Yakshina}}, \bibinfo {author} {\bibfnamefont {D.~B.}\ \bibnamefont {Tretyakov}}, \bibinfo {author} {\bibfnamefont {V.~M.}\ \bibnamefont {Entin}}, \bibinfo {author} {\bibfnamefont {I.~I.}\ \bibnamefont {Ryabtsev}}, \bibinfo {author} {\bibfnamefont {P.}~\bibnamefont {Cheinet}}, \bibinfo {author} {\bibfnamefont {P.}~\bibnamefont {Pillet}},\ and\ \bibinfo {author} {\bibfnamefont {M.}~\bibnamefont {Saffman}},\ }\bibfield  {title} {\bibinfo {title} {Fast three-qubit toffoli quantum gate based on three-body f\"orster resonances in rydberg atoms},\ }\href {https://doi.org/10.1103/PhysRevA.98.042704} {\bibfield  {journal} {\bibinfo  {journal} {Phys. Rev. A}\ }\textbf {\bibinfo {volume} {98}},\ \bibinfo {pages} {042704} (\bibinfo {year} {2018})}\BibitemShut {NoStop}%
\bibitem [{\citenamefont {Levine}\ \emph {et~al.}(2019)\citenamefont {Levine}, \citenamefont {Keesling}, \citenamefont {Semeghini}, \citenamefont {Omran}, \citenamefont {Wang}, \citenamefont {Ebadi}, \citenamefont {Bernien}, \citenamefont {Greiner}, \citenamefont {Vuleti\ifmmode~\acute{c}\else \'{c}\fi{}}, \citenamefont {Pichler},\ and\ \citenamefont {Lukin}}]{PhysRevLett.123.170503}%
  \BibitemOpen
  \bibfield  {author} {\bibinfo {author} {\bibfnamefont {H.}~\bibnamefont {Levine}}, \bibinfo {author} {\bibfnamefont {A.}~\bibnamefont {Keesling}}, \bibinfo {author} {\bibfnamefont {G.}~\bibnamefont {Semeghini}}, \bibinfo {author} {\bibfnamefont {A.}~\bibnamefont {Omran}}, \bibinfo {author} {\bibfnamefont {T.~T.}\ \bibnamefont {Wang}}, \bibinfo {author} {\bibfnamefont {S.}~\bibnamefont {Ebadi}}, \bibinfo {author} {\bibfnamefont {H.}~\bibnamefont {Bernien}}, \bibinfo {author} {\bibfnamefont {M.}~\bibnamefont {Greiner}}, \bibinfo {author} {\bibfnamefont {V.}~\bibnamefont {Vuleti\ifmmode~\acute{c}\else \'{c}\fi{}}}, \bibinfo {author} {\bibfnamefont {H.}~\bibnamefont {Pichler}},\ and\ \bibinfo {author} {\bibfnamefont {M.~D.}\ \bibnamefont {Lukin}},\ }\bibfield  {title} {\bibinfo {title} {Parallel implementation of high-fidelity multiqubit gates with neutral atoms},\ }\href {https://doi.org/10.1103/PhysRevLett.123.170503} {\bibfield  {journal} {\bibinfo  {journal} {Phys. Rev. Lett.}\ }\textbf {\bibinfo {volume}
  {123}},\ \bibinfo {pages} {170503} (\bibinfo {year} {2019})}\BibitemShut {NoStop}%
\bibitem [{\citenamefont {Yin}\ \emph {et~al.}(2020)\citenamefont {Yin}, \citenamefont {Li}, \citenamefont {Wang},\ and\ \citenamefont {Shao}}]{Yin:20}%
  \BibitemOpen
  \bibfield  {author} {\bibinfo {author} {\bibfnamefont {H.-D.}\ \bibnamefont {Yin}}, \bibinfo {author} {\bibfnamefont {X.-X.}\ \bibnamefont {Li}}, \bibinfo {author} {\bibfnamefont {G.-C.}\ \bibnamefont {Wang}},\ and\ \bibinfo {author} {\bibfnamefont {X.-Q.}\ \bibnamefont {Shao}},\ }\bibfield  {title} {\bibinfo {title} {One-step implementation of toffoli gate for neutral atoms based on unconventional rydberg pumping},\ }\href {https://doi.org/10.1364/OE.410158} {\bibfield  {journal} {\bibinfo  {journal} {Opt. Express}\ }\textbf {\bibinfo {volume} {28}},\ \bibinfo {pages} {35576} (\bibinfo {year} {2020})}\BibitemShut {NoStop}%
\bibitem [{\citenamefont {Yin}\ and\ \citenamefont {Shao}(2021)}]{Yin20212541}%
  \BibitemOpen
  \bibfield  {author} {\bibinfo {author} {\bibfnamefont {H.-D.}\ \bibnamefont {Yin}}\ and\ \bibinfo {author} {\bibfnamefont {X.-Q.}\ \bibnamefont {Shao}},\ }\bibfield  {title} {\bibinfo {title} {Gaussian soft control-based quantum fan-out gate in ground-state manifolds of neutral atoms},\ }\href {https://doi.org/10.1364/OL.424469} {\bibfield  {journal} {\bibinfo  {journal} {Optics Letters}\ }\textbf {\bibinfo {volume} {46}},\ \bibinfo {pages} {2541 – 2544} (\bibinfo {year} {2021})}\BibitemShut {NoStop}%
\bibitem [{\citenamefont {Wang}\ \emph {et~al.}(2025)\citenamefont {Wang}, \citenamefont {Wang}, \citenamefont {Dong}, \citenamefont {Xiu},\ and\ \citenamefont {Ji}}]{WANG2025112812}%
  \BibitemOpen
  \bibfield  {author} {\bibinfo {author} {\bibfnamefont {J.}~\bibnamefont {Wang}}, \bibinfo {author} {\bibfnamefont {J.}~\bibnamefont {Wang}}, \bibinfo {author} {\bibfnamefont {L.}~\bibnamefont {Dong}}, \bibinfo {author} {\bibfnamefont {X.}~\bibnamefont {Xiu}},\ and\ \bibinfo {author} {\bibfnamefont {Y.}~\bibnamefont {Ji}},\ }\bibfield  {title} {\bibinfo {title} {Fast generation of quantum logic gate with rydberg superatoms},\ }\href {https://doi.org/https://doi.org/10.1016/j.optlastec.2025.112812} {\bibfield  {journal} {\bibinfo  {journal} {Optics \& Laser Technology}\ }\textbf {\bibinfo {volume} {187}},\ \bibinfo {pages} {112812} (\bibinfo {year} {2025})}\BibitemShut {NoStop}%
\bibitem [{\citenamefont {Wallquist}\ \emph {et~al.}(2009)\citenamefont {Wallquist}, \citenamefont {Hammerer}, \citenamefont {Rabl}, \citenamefont {Lukin},\ and\ \citenamefont {Zoller}}]{Wallquist:2009mwr}%
  \BibitemOpen
  \bibfield  {author} {\bibinfo {author} {\bibfnamefont {M.}~\bibnamefont {Wallquist}}, \bibinfo {author} {\bibfnamefont {K.}~\bibnamefont {Hammerer}}, \bibinfo {author} {\bibfnamefont {P.}~\bibnamefont {Rabl}}, \bibinfo {author} {\bibfnamefont {M.}~\bibnamefont {Lukin}},\ and\ \bibinfo {author} {\bibfnamefont {P.}~\bibnamefont {Zoller}},\ }\bibfield  {title} {\bibinfo {title} {{Hybrid quantum devices and quantum engineering}},\ }\href {https://doi.org/10.1088/0031-8949/2009/T137/014001} {\bibfield  {journal} {\bibinfo  {journal} {Phys. Scripta}\ }\textbf {\bibinfo {volume} {2009}},\ \bibinfo {pages} {014001} (\bibinfo {year} {2009})}\BibitemShut {NoStop}%
\bibitem [{\citenamefont {Morton}\ and\ \citenamefont {Lovett}(2011)}]{annurev:/content/journals/10.1146/annurev-conmatphys-062910-140514}%
  \BibitemOpen
  \bibfield  {author} {\bibinfo {author} {\bibfnamefont {J.~J.}\ \bibnamefont {Morton}}\ and\ \bibinfo {author} {\bibfnamefont {B.~W.}\ \bibnamefont {Lovett}},\ }\bibfield  {title} {\bibinfo {title} {Hybrid solid-state qubits: The powerful role of electron spins},\ }\href {https://doi.org/https://doi.org/10.1146/annurev-conmatphys-062910-140514} {\bibfield  {journal} {\bibinfo  {journal} {Annual Review of Condensed Matter Physics}\ }\textbf {\bibinfo {volume} {2}},\ \bibinfo {pages} {189} (\bibinfo {year} {2011})}\BibitemShut {NoStop}%
\bibitem [{\citenamefont {Xiang}\ \emph {et~al.}(2013)\citenamefont {Xiang}, \citenamefont {Ashhab}, \citenamefont {You},\ and\ \citenamefont {Nori}}]{RevModPhys.85.623}%
  \BibitemOpen
  \bibfield  {author} {\bibinfo {author} {\bibfnamefont {Z.-L.}\ \bibnamefont {Xiang}}, \bibinfo {author} {\bibfnamefont {S.}~\bibnamefont {Ashhab}}, \bibinfo {author} {\bibfnamefont {J.~Q.}\ \bibnamefont {You}},\ and\ \bibinfo {author} {\bibfnamefont {F.}~\bibnamefont {Nori}},\ }\bibfield  {title} {\bibinfo {title} {Hybrid quantum circuits: Superconducting circuits interacting with other quantum systems},\ }\href {https://doi.org/10.1103/RevModPhys.85.623} {\bibfield  {journal} {\bibinfo  {journal} {Rev. Mod. Phys.}\ }\textbf {\bibinfo {volume} {85}},\ \bibinfo {pages} {623} (\bibinfo {year} {2013})}\BibitemShut {NoStop}%
\bibitem [{\citenamefont {Daniilidis}\ and\ \citenamefont {Häffner}(2013)}]{annurev:/content/journals/10.1146/annurev-conmatphys-030212-184253}%
  \BibitemOpen
  \bibfield  {author} {\bibinfo {author} {\bibfnamefont {N.}~\bibnamefont {Daniilidis}}\ and\ \bibinfo {author} {\bibfnamefont {H.}~\bibnamefont {Häffner}},\ }\bibfield  {title} {\bibinfo {title} {Quantum interfaces between atomic and solid-state systems},\ }\href {https://doi.org/https://doi.org/10.1146/annurev-conmatphys-030212-184253} {\bibfield  {journal} {\bibinfo  {journal} {Annual Review of Condensed Matter Physics}\ }\textbf {\bibinfo {volume} {4}},\ \bibinfo {pages} {83} (\bibinfo {year} {2013})}\BibitemShut {NoStop}%
\bibitem [{\citenamefont {Kurizki}\ \emph {et~al.}(2015)\citenamefont {Kurizki}, \citenamefont {Bertet}, \citenamefont {Kubo}, \citenamefont {Mølmer}, \citenamefont {Petrosyan}, \citenamefont {Rabl},\ and\ \citenamefont {Schmiedmayer}}]{doi:10.1073/pnas.1419326112}%
  \BibitemOpen
  \bibfield  {author} {\bibinfo {author} {\bibfnamefont {G.}~\bibnamefont {Kurizki}}, \bibinfo {author} {\bibfnamefont {P.}~\bibnamefont {Bertet}}, \bibinfo {author} {\bibfnamefont {Y.}~\bibnamefont {Kubo}}, \bibinfo {author} {\bibfnamefont {K.}~\bibnamefont {Mølmer}}, \bibinfo {author} {\bibfnamefont {D.}~\bibnamefont {Petrosyan}}, \bibinfo {author} {\bibfnamefont {P.}~\bibnamefont {Rabl}},\ and\ \bibinfo {author} {\bibfnamefont {J.}~\bibnamefont {Schmiedmayer}},\ }\bibfield  {title} {\bibinfo {title} {Quantum technologies with hybrid systems},\ }\href {https://doi.org/10.1073/pnas.1419326112} {\bibfield  {journal} {\bibinfo  {journal} {Proceedings of the National Academy of Sciences}\ }\textbf {\bibinfo {volume} {112}},\ \bibinfo {pages} {3866} (\bibinfo {year} {2015})}\BibitemShut {NoStop}%
\bibitem [{\citenamefont {Beterov}\ \emph {et~al.}(2009)\citenamefont {Beterov}, \citenamefont {Ryabtsev}, \citenamefont {Tretyakov},\ and\ \citenamefont {Entin}}]{PhysRevA.79.052504}%
  \BibitemOpen
  \bibfield  {author} {\bibinfo {author} {\bibfnamefont {I.~I.}\ \bibnamefont {Beterov}}, \bibinfo {author} {\bibfnamefont {I.~I.}\ \bibnamefont {Ryabtsev}}, \bibinfo {author} {\bibfnamefont {D.~B.}\ \bibnamefont {Tretyakov}},\ and\ \bibinfo {author} {\bibfnamefont {V.~M.}\ \bibnamefont {Entin}},\ }\bibfield  {title} {\bibinfo {title} {Quasiclassical calculations of blackbody-radiation-induced depopulation rates and effective lifetimes of rydberg $ns$, $np$, and $nd$ alkali-metal atoms with $n\ensuremath{\le}80$},\ }\href {https://doi.org/10.1103/PhysRevA.79.052504} {\bibfield  {journal} {\bibinfo  {journal} {Phys. Rev. A}\ }\textbf {\bibinfo {volume} {79}},\ \bibinfo {pages} {052504} (\bibinfo {year} {2009})}\BibitemShut {NoStop}%
\bibitem [{\citenamefont {Flannery}\ \emph {et~al.}(2005)\citenamefont {Flannery}, \citenamefont {Vrinceanu},\ and\ \citenamefont {Ostrovsky}}]{Flannery_2005}%
  \BibitemOpen
  \bibfield  {author} {\bibinfo {author} {\bibfnamefont {M.~R.}\ \bibnamefont {Flannery}}, \bibinfo {author} {\bibfnamefont {D.}~\bibnamefont {Vrinceanu}},\ and\ \bibinfo {author} {\bibfnamefont {V.~N.}\ \bibnamefont {Ostrovsky}},\ }\bibfield  {title} {\bibinfo {title} {Long-range interaction between polar rydberg atoms},\ }\href {https://doi.org/10.1088/0953-4075/38/2/020} {\bibfield  {journal} {\bibinfo  {journal} {Journal of Physics B: Atomic, Molecular and Optical Physics}\ }\textbf {\bibinfo {volume} {38}},\ \bibinfo {pages} {S279} (\bibinfo {year} {2005})}\BibitemShut {NoStop}%
\bibitem [{\citenamefont {Browaeys}\ \emph {et~al.}(2016)\citenamefont {Browaeys}, \citenamefont {Barredo},\ and\ \citenamefont {Lahaye}}]{Browaeys_2016}%
  \BibitemOpen
  \bibfield  {author} {\bibinfo {author} {\bibfnamefont {A.}~\bibnamefont {Browaeys}}, \bibinfo {author} {\bibfnamefont {D.}~\bibnamefont {Barredo}},\ and\ \bibinfo {author} {\bibfnamefont {T.}~\bibnamefont {Lahaye}},\ }\bibfield  {title} {\bibinfo {title} {Experimental investigations of dipole–dipole interactions between a few rydberg atoms},\ }\href {https://doi.org/10.1088/0953-4075/49/15/152001} {\bibfield  {journal} {\bibinfo  {journal} {Journal of Physics B: Atomic, Molecular and Optical Physics}\ }\textbf {\bibinfo {volume} {49}},\ \bibinfo {pages} {152001} (\bibinfo {year} {2016})}\BibitemShut {NoStop}%
\bibitem [{\citenamefont {Saffman}(2016)}]{Saffman_2016}%
  \BibitemOpen
  \bibfield  {author} {\bibinfo {author} {\bibfnamefont {M.}~\bibnamefont {Saffman}},\ }\bibfield  {title} {\bibinfo {title} {Quantum computing with atomic qubits and rydberg interactions: progress and challenges},\ }\href {https://doi.org/10.1088/0953-4075/49/20/202001} {\bibfield  {journal} {\bibinfo  {journal} {Journal of Physics B: Atomic, Molecular and Optical Physics}\ }\textbf {\bibinfo {volume} {49}},\ \bibinfo {pages} {202001} (\bibinfo {year} {2016})}\BibitemShut {NoStop}%
\bibitem [{\citenamefont {Zhang}\ and\ \citenamefont {Ye}(2016)}]{10.1093/nsr/nww013}%
  \BibitemOpen
  \bibfield  {author} {\bibinfo {author} {\bibfnamefont {X.}~\bibnamefont {Zhang}}\ and\ \bibinfo {author} {\bibfnamefont {J.}~\bibnamefont {Ye}},\ }\bibfield  {title} {\bibinfo {title} {Precision measurement and frequency metrology with ultracold atoms},\ }\href {https://doi.org/10.1093/nsr/nww013} {\bibfield  {journal} {\bibinfo  {journal} {National Science Review}\ }\textbf {\bibinfo {volume} {3}},\ \bibinfo {pages} {189} (\bibinfo {year} {2016})}\BibitemShut {NoStop}%
\bibitem [{\citenamefont {Gross}\ and\ \citenamefont {Bloch}(2017)}]{doi:10.1126/science.aal3837}%
  \BibitemOpen
  \bibfield  {author} {\bibinfo {author} {\bibfnamefont {C.}~\bibnamefont {Gross}}\ and\ \bibinfo {author} {\bibfnamefont {I.}~\bibnamefont {Bloch}},\ }\bibfield  {title} {\bibinfo {title} {Quantum simulations with ultracold atoms in optical lattices},\ }\href {https://doi.org/10.1126/science.aal3837} {\bibfield  {journal} {\bibinfo  {journal} {Science}\ }\textbf {\bibinfo {volume} {357}},\ \bibinfo {pages} {995} (\bibinfo {year} {2017})}\BibitemShut {NoStop}%
\bibitem [{\citenamefont {Picken}\ \emph {et~al.}(2018)\citenamefont {Picken}, \citenamefont {Legaie}, \citenamefont {McDonnell},\ and\ \citenamefont {Pritchard}}]{Picken_2019}%
  \BibitemOpen
  \bibfield  {author} {\bibinfo {author} {\bibfnamefont {C.~J.}\ \bibnamefont {Picken}}, \bibinfo {author} {\bibfnamefont {R.}~\bibnamefont {Legaie}}, \bibinfo {author} {\bibfnamefont {K.}~\bibnamefont {McDonnell}},\ and\ \bibinfo {author} {\bibfnamefont {J.~D.}\ \bibnamefont {Pritchard}},\ }\bibfield  {title} {\bibinfo {title} {Entanglement of neutral-atom qubits with long ground-rydberg coherence times},\ }\href {https://doi.org/10.1088/2058-9565/aaf019} {\bibfield  {journal} {\bibinfo  {journal} {Quantum Science and Technology}\ }\textbf {\bibinfo {volume} {4}},\ \bibinfo {pages} {015011} (\bibinfo {year} {2018})}\BibitemShut {NoStop}%
\bibitem [{\citenamefont {Idziaszek}\ \emph {et~al.}(2007)\citenamefont {Idziaszek}, \citenamefont {Calarco},\ and\ \citenamefont {Zoller}}]{PhysRevA.76.033409}%
  \BibitemOpen
  \bibfield  {author} {\bibinfo {author} {\bibfnamefont {Z.}~\bibnamefont {Idziaszek}}, \bibinfo {author} {\bibfnamefont {T.}~\bibnamefont {Calarco}},\ and\ \bibinfo {author} {\bibfnamefont {P.}~\bibnamefont {Zoller}},\ }\bibfield  {title} {\bibinfo {title} {Controlled collisions of a single atom and an ion guided by movable trapping potentials},\ }\href {https://doi.org/10.1103/PhysRevA.76.033409} {\bibfield  {journal} {\bibinfo  {journal} {Phys. Rev. A}\ }\textbf {\bibinfo {volume} {76}},\ \bibinfo {pages} {033409} (\bibinfo {year} {2007})}\BibitemShut {NoStop}%
\bibitem [{\citenamefont {Doerk}\ \emph {et~al.}(2010)\citenamefont {Doerk}, \citenamefont {Idziaszek},\ and\ \citenamefont {Calarco}}]{PhysRevA.81.012708}%
  \BibitemOpen
  \bibfield  {author} {\bibinfo {author} {\bibfnamefont {H.}~\bibnamefont {Doerk}}, \bibinfo {author} {\bibfnamefont {Z.}~\bibnamefont {Idziaszek}},\ and\ \bibinfo {author} {\bibfnamefont {T.}~\bibnamefont {Calarco}},\ }\bibfield  {title} {\bibinfo {title} {Atom-ion quantum gate},\ }\href {https://doi.org/10.1103/PhysRevA.81.012708} {\bibfield  {journal} {\bibinfo  {journal} {Phys. Rev. A}\ }\textbf {\bibinfo {volume} {81}},\ \bibinfo {pages} {012708} (\bibinfo {year} {2010})}\BibitemShut {NoStop}%
\bibitem [{\citenamefont {Secker}\ \emph {et~al.}(2016)\citenamefont {Secker}, \citenamefont {Gerritsma}, \citenamefont {Glaetzle},\ and\ \citenamefont {Negretti}}]{PhysRevA.94.013420}%
  \BibitemOpen
  \bibfield  {author} {\bibinfo {author} {\bibfnamefont {T.}~\bibnamefont {Secker}}, \bibinfo {author} {\bibfnamefont {R.}~\bibnamefont {Gerritsma}}, \bibinfo {author} {\bibfnamefont {A.~W.}\ \bibnamefont {Glaetzle}},\ and\ \bibinfo {author} {\bibfnamefont {A.}~\bibnamefont {Negretti}},\ }\bibfield  {title} {\bibinfo {title} {Controlled long-range interactions between rydberg atoms and ions},\ }\href {https://doi.org/10.1103/PhysRevA.94.013420} {\bibfield  {journal} {\bibinfo  {journal} {Phys. Rev. A}\ }\textbf {\bibinfo {volume} {94}},\ \bibinfo {pages} {013420} (\bibinfo {year} {2016})}\BibitemShut {NoStop}%
\bibitem [{\citenamefont {Tomza}\ \emph {et~al.}(2019)\citenamefont {Tomza}, \citenamefont {Jachymski}, \citenamefont {Gerritsma}, \citenamefont {Negretti}, \citenamefont {Calarco}, \citenamefont {Idziaszek},\ and\ \citenamefont {Julienne}}]{RevModPhys.91.035001}%
  \BibitemOpen
  \bibfield  {author} {\bibinfo {author} {\bibfnamefont {M.}~\bibnamefont {Tomza}}, \bibinfo {author} {\bibfnamefont {K.}~\bibnamefont {Jachymski}}, \bibinfo {author} {\bibfnamefont {R.}~\bibnamefont {Gerritsma}}, \bibinfo {author} {\bibfnamefont {A.}~\bibnamefont {Negretti}}, \bibinfo {author} {\bibfnamefont {T.}~\bibnamefont {Calarco}}, \bibinfo {author} {\bibfnamefont {Z.}~\bibnamefont {Idziaszek}},\ and\ \bibinfo {author} {\bibfnamefont {P.~S.}\ \bibnamefont {Julienne}},\ }\bibfield  {title} {\bibinfo {title} {Cold hybrid ion-atom systems},\ }\href {https://doi.org/10.1103/RevModPhys.91.035001} {\bibfield  {journal} {\bibinfo  {journal} {Rev. Mod. Phys.}\ }\textbf {\bibinfo {volume} {91}},\ \bibinfo {pages} {035001} (\bibinfo {year} {2019})}\BibitemShut {NoStop}%
\bibitem [{\citenamefont {Shaposhnikov}\ and\ \citenamefont {Fedichkin}(2023)}]{Shaposhnikov2023}%
  \BibitemOpen
  \bibfield  {author} {\bibinfo {author} {\bibfnamefont {D.}~\bibnamefont {Shaposhnikov}}\ and\ \bibinfo {author} {\bibfnamefont {L.}~\bibnamefont {Fedichkin}},\ }\bibfield  {title} {\bibinfo {title} {Hybrid atom-ion quantum gate engineering},\ }\href {https://doi.org/10.1134/S1063739723600437} {\bibfield  {journal} {\bibinfo  {journal} {Russian Microelectronics}\ }\textbf {\bibinfo {volume} {52}},\ \bibinfo {pages} {S369} (\bibinfo {year} {2023})}\BibitemShut {NoStop}%
\bibitem [{\citenamefont {Duan}\ and\ \citenamefont {Kimble}(2004)}]{PhysRevLett.92.127902}%
  \BibitemOpen
  \bibfield  {author} {\bibinfo {author} {\bibfnamefont {L.-M.}\ \bibnamefont {Duan}}\ and\ \bibinfo {author} {\bibfnamefont {H.~J.}\ \bibnamefont {Kimble}},\ }\bibfield  {title} {\bibinfo {title} {Scalable photonic quantum computation through cavity-assisted interactions},\ }\href {https://doi.org/10.1103/PhysRevLett.92.127902} {\bibfield  {journal} {\bibinfo  {journal} {Phys. Rev. Lett.}\ }\textbf {\bibinfo {volume} {92}},\ \bibinfo {pages} {127902} (\bibinfo {year} {2004})}\BibitemShut {NoStop}%
\bibitem [{\citenamefont {Reiserer}\ \emph {et~al.}(2013)\citenamefont {Reiserer}, \citenamefont {Ritter},\ and\ \citenamefont {Rempe}}]{doi:10.1126/science.1246164}%
  \BibitemOpen
  \bibfield  {author} {\bibinfo {author} {\bibfnamefont {A.}~\bibnamefont {Reiserer}}, \bibinfo {author} {\bibfnamefont {S.}~\bibnamefont {Ritter}},\ and\ \bibinfo {author} {\bibfnamefont {G.}~\bibnamefont {Rempe}},\ }\bibfield  {title} {\bibinfo {title} {Nondestructive detection of an optical photon},\ }\href {https://doi.org/10.1126/science.1246164} {\bibfield  {journal} {\bibinfo  {journal} {Science}\ }\textbf {\bibinfo {volume} {342}},\ \bibinfo {pages} {1349} (\bibinfo {year} {2013})}\BibitemShut {NoStop}%
\bibitem [{\citenamefont {Reiserer}\ \emph {et~al.}(2014)\citenamefont {Reiserer}, \citenamefont {Kalb}, \citenamefont {Rempe},\ and\ \citenamefont {Ritter}}]{Reiserer2014}%
  \BibitemOpen
  \bibfield  {author} {\bibinfo {author} {\bibfnamefont {A.}~\bibnamefont {Reiserer}}, \bibinfo {author} {\bibfnamefont {N.}~\bibnamefont {Kalb}}, \bibinfo {author} {\bibfnamefont {G.}~\bibnamefont {Rempe}},\ and\ \bibinfo {author} {\bibfnamefont {S.}~\bibnamefont {Ritter}},\ }\bibfield  {title} {\bibinfo {title} {A quantum gate between a flying optical photon and a single trapped atom},\ }\href {https://doi.org/10.1038/nature13177} {\bibfield  {journal} {\bibinfo  {journal} {Nature}\ }\textbf {\bibinfo {volume} {508}},\ \bibinfo {pages} {237} (\bibinfo {year} {2014})}\BibitemShut {NoStop}%
\bibitem [{\citenamefont {Langenfeld}\ \emph {et~al.}(2021)\citenamefont {Langenfeld}, \citenamefont {Welte}, \citenamefont {Hartung}, \citenamefont {Daiss}, \citenamefont {Thomas}, \citenamefont {Morin}, \citenamefont {Distante},\ and\ \citenamefont {Rempe}}]{PhysRevLett.126.130502}%
  \BibitemOpen
  \bibfield  {author} {\bibinfo {author} {\bibfnamefont {S.}~\bibnamefont {Langenfeld}}, \bibinfo {author} {\bibfnamefont {S.}~\bibnamefont {Welte}}, \bibinfo {author} {\bibfnamefont {L.}~\bibnamefont {Hartung}}, \bibinfo {author} {\bibfnamefont {S.}~\bibnamefont {Daiss}}, \bibinfo {author} {\bibfnamefont {P.}~\bibnamefont {Thomas}}, \bibinfo {author} {\bibfnamefont {O.}~\bibnamefont {Morin}}, \bibinfo {author} {\bibfnamefont {E.}~\bibnamefont {Distante}},\ and\ \bibinfo {author} {\bibfnamefont {G.}~\bibnamefont {Rempe}},\ }\bibfield  {title} {\bibinfo {title} {Quantum teleportation between remote qubit memories with only a single photon as a resource},\ }\href {https://doi.org/10.1103/PhysRevLett.126.130502} {\bibfield  {journal} {\bibinfo  {journal} {Phys. Rev. Lett.}\ }\textbf {\bibinfo {volume} {126}},\ \bibinfo {pages} {130502} (\bibinfo {year} {2021})}\BibitemShut {NoStop}%
\bibitem [{\citenamefont {Nagib}\ \emph {et~al.}(2024)\citenamefont {Nagib}, \citenamefont {Huft}, \citenamefont {Safari},\ and\ \citenamefont {Saffman}}]{PhysRevA.109.032602}%
  \BibitemOpen
  \bibfield  {author} {\bibinfo {author} {\bibfnamefont {O.}~\bibnamefont {Nagib}}, \bibinfo {author} {\bibfnamefont {P.}~\bibnamefont {Huft}}, \bibinfo {author} {\bibfnamefont {A.}~\bibnamefont {Safari}},\ and\ \bibinfo {author} {\bibfnamefont {M.}~\bibnamefont {Saffman}},\ }\bibfield  {title} {\bibinfo {title} {Robust atom-photon gate for quantum information processing},\ }\href {https://doi.org/10.1103/PhysRevA.109.032602} {\bibfield  {journal} {\bibinfo  {journal} {Phys. Rev. A}\ }\textbf {\bibinfo {volume} {109}},\ \bibinfo {pages} {032602} (\bibinfo {year} {2024})}\BibitemShut {NoStop}%
\bibitem [{\citenamefont {Sawant}\ \emph {et~al.}(2020)\citenamefont {Sawant}, \citenamefont {Blackmore}, \citenamefont {Gregory}, \citenamefont {Mur-Petit}, \citenamefont {Jaksch}, \citenamefont {Aldegunde}, \citenamefont {Hutson}, \citenamefont {Tarbutt},\ and\ \citenamefont {Cornish}}]{Sawant_2020}%
  \BibitemOpen
  \bibfield  {author} {\bibinfo {author} {\bibfnamefont {R.}~\bibnamefont {Sawant}}, \bibinfo {author} {\bibfnamefont {J.~A.}\ \bibnamefont {Blackmore}}, \bibinfo {author} {\bibfnamefont {P.~D.}\ \bibnamefont {Gregory}}, \bibinfo {author} {\bibfnamefont {J.}~\bibnamefont {Mur-Petit}}, \bibinfo {author} {\bibfnamefont {D.}~\bibnamefont {Jaksch}}, \bibinfo {author} {\bibfnamefont {J.}~\bibnamefont {Aldegunde}}, \bibinfo {author} {\bibfnamefont {J.~M.}\ \bibnamefont {Hutson}}, \bibinfo {author} {\bibfnamefont {M.~R.}\ \bibnamefont {Tarbutt}},\ and\ \bibinfo {author} {\bibfnamefont {S.~L.}\ \bibnamefont {Cornish}},\ }\bibfield  {title} {\bibinfo {title} {Ultracold polar molecules as qudits},\ }\href {https://doi.org/10.1088/1367-2630/ab60f4} {\bibfield  {journal} {\bibinfo  {journal} {New Journal of Physics}\ }\textbf {\bibinfo {volume} {22}},\ \bibinfo {pages} {013027} (\bibinfo {year} {2020})}\BibitemShut {NoStop}%
\bibitem [{\citenamefont {Gregory}\ \emph {et~al.}(2021)\citenamefont {Gregory}, \citenamefont {Blackmore}, \citenamefont {Bromley}, \citenamefont {Hutson},\ and\ \citenamefont {Cornish}}]{Gregory2021}%
  \BibitemOpen
  \bibfield  {author} {\bibinfo {author} {\bibfnamefont {P.~D.}\ \bibnamefont {Gregory}}, \bibinfo {author} {\bibfnamefont {J.~A.}\ \bibnamefont {Blackmore}}, \bibinfo {author} {\bibfnamefont {S.~L.}\ \bibnamefont {Bromley}}, \bibinfo {author} {\bibfnamefont {J.~M.}\ \bibnamefont {Hutson}},\ and\ \bibinfo {author} {\bibfnamefont {S.~L.}\ \bibnamefont {Cornish}},\ }\bibfield  {title} {\bibinfo {title} {Robust storage qubits in ultracold polar molecules},\ }\href {https://doi.org/10.1038/s41567-021-01328-7} {\bibfield  {journal} {\bibinfo  {journal} {Nature Physics}\ }\textbf {\bibinfo {volume} {17}},\ \bibinfo {pages} {1149} (\bibinfo {year} {2021})}\BibitemShut {NoStop}%
\bibitem [{\citenamefont {Burchesky}\ \emph {et~al.}(2021)\citenamefont {Burchesky}, \citenamefont {Anderegg}, \citenamefont {Bao}, \citenamefont {Yu}, \citenamefont {Chae}, \citenamefont {Ketterle}, \citenamefont {Ni},\ and\ \citenamefont {Doyle}}]{PhysRevLett.127.123202}%
  \BibitemOpen
  \bibfield  {author} {\bibinfo {author} {\bibfnamefont {S.}~\bibnamefont {Burchesky}}, \bibinfo {author} {\bibfnamefont {L.}~\bibnamefont {Anderegg}}, \bibinfo {author} {\bibfnamefont {Y.}~\bibnamefont {Bao}}, \bibinfo {author} {\bibfnamefont {S.~S.}\ \bibnamefont {Yu}}, \bibinfo {author} {\bibfnamefont {E.}~\bibnamefont {Chae}}, \bibinfo {author} {\bibfnamefont {W.}~\bibnamefont {Ketterle}}, \bibinfo {author} {\bibfnamefont {K.-K.}\ \bibnamefont {Ni}},\ and\ \bibinfo {author} {\bibfnamefont {J.~M.}\ \bibnamefont {Doyle}},\ }\bibfield  {title} {\bibinfo {title} {Rotational coherence times of polar molecules in optical tweezers},\ }\href {https://doi.org/10.1103/PhysRevLett.127.123202} {\bibfield  {journal} {\bibinfo  {journal} {Phys. Rev. Lett.}\ }\textbf {\bibinfo {volume} {127}},\ \bibinfo {pages} {123202} (\bibinfo {year} {2021})}\BibitemShut {NoStop}%
\bibitem [{\citenamefont {Yan}\ \emph {et~al.}(2013)\citenamefont {Yan}, \citenamefont {Moses}, \citenamefont {Gadway}, \citenamefont {Covey}, \citenamefont {Hazzard}, \citenamefont {Rey}, \citenamefont {Jin},\ and\ \citenamefont {Ye}}]{Yan2013}%
  \BibitemOpen
  \bibfield  {author} {\bibinfo {author} {\bibfnamefont {B.}~\bibnamefont {Yan}}, \bibinfo {author} {\bibfnamefont {S.~A.}\ \bibnamefont {Moses}}, \bibinfo {author} {\bibfnamefont {B.}~\bibnamefont {Gadway}}, \bibinfo {author} {\bibfnamefont {J.~P.}\ \bibnamefont {Covey}}, \bibinfo {author} {\bibfnamefont {K.~R.~A.}\ \bibnamefont {Hazzard}}, \bibinfo {author} {\bibfnamefont {A.~M.}\ \bibnamefont {Rey}}, \bibinfo {author} {\bibfnamefont {D.~S.}\ \bibnamefont {Jin}},\ and\ \bibinfo {author} {\bibfnamefont {J.}~\bibnamefont {Ye}},\ }\bibfield  {title} {\bibinfo {title} {Observation of dipolar spin-exchange interactions with lattice-confined polar molecules},\ }\href {https://doi.org/10.1038/nature12483} {\bibfield  {journal} {\bibinfo  {journal} {Nature}\ }\textbf {\bibinfo {volume} {501}},\ \bibinfo {pages} {521} (\bibinfo {year} {2013})}\BibitemShut {NoStop}%
\bibitem [{\citenamefont {Tobias}\ \emph {et~al.}(2022)\citenamefont {Tobias}, \citenamefont {Matsuda}, \citenamefont {Li}, \citenamefont {Miller}, \citenamefont {Carroll}, \citenamefont {Bilitewski}, \citenamefont {Rey},\ and\ \citenamefont {Ye}}]{doi:10.1126/science.abn8525}%
  \BibitemOpen
  \bibfield  {author} {\bibinfo {author} {\bibfnamefont {W.~G.}\ \bibnamefont {Tobias}}, \bibinfo {author} {\bibfnamefont {K.}~\bibnamefont {Matsuda}}, \bibinfo {author} {\bibfnamefont {J.-R.}\ \bibnamefont {Li}}, \bibinfo {author} {\bibfnamefont {C.}~\bibnamefont {Miller}}, \bibinfo {author} {\bibfnamefont {A.~N.}\ \bibnamefont {Carroll}}, \bibinfo {author} {\bibfnamefont {T.}~\bibnamefont {Bilitewski}}, \bibinfo {author} {\bibfnamefont {A.~M.}\ \bibnamefont {Rey}},\ and\ \bibinfo {author} {\bibfnamefont {J.}~\bibnamefont {Ye}},\ }\bibfield  {title} {\bibinfo {title} {Reactions between layer-resolved molecules mediated by dipolar spin exchange},\ }\href {https://doi.org/10.1126/science.abn8525} {\bibfield  {journal} {\bibinfo  {journal} {Science}\ }\textbf {\bibinfo {volume} {375}},\ \bibinfo {pages} {1299} (\bibinfo {year} {2022})}\BibitemShut {NoStop}%
\bibitem [{\citenamefont {Wang}\ \emph {et~al.}(2022)\citenamefont {Wang}, \citenamefont {Williams}, \citenamefont {Picard}, \citenamefont {Yao},\ and\ \citenamefont {Ni}}]{PRXQuantum.3.030339}%
  \BibitemOpen
  \bibfield  {author} {\bibinfo {author} {\bibfnamefont {K.}~\bibnamefont {Wang}}, \bibinfo {author} {\bibfnamefont {C.~P.}\ \bibnamefont {Williams}}, \bibinfo {author} {\bibfnamefont {L.~R.}\ \bibnamefont {Picard}}, \bibinfo {author} {\bibfnamefont {N.~Y.}\ \bibnamefont {Yao}},\ and\ \bibinfo {author} {\bibfnamefont {K.-K.}\ \bibnamefont {Ni}},\ }\bibfield  {title} {\bibinfo {title} {Enriching the quantum toolbox of ultracold molecules with rydberg atoms},\ }\href {https://doi.org/10.1103/PRXQuantum.3.030339} {\bibfield  {journal} {\bibinfo  {journal} {PRX Quantum}\ }\textbf {\bibinfo {volume} {3}},\ \bibinfo {pages} {030339} (\bibinfo {year} {2022})}\BibitemShut {NoStop}%
\bibitem [{\citenamefont {Zhang}\ and\ \citenamefont {Tarbutt}(2022)}]{PRXQuantum.3.030340}%
  \BibitemOpen
  \bibfield  {author} {\bibinfo {author} {\bibfnamefont {C.}~\bibnamefont {Zhang}}\ and\ \bibinfo {author} {\bibfnamefont {M.}~\bibnamefont {Tarbutt}},\ }\bibfield  {title} {\bibinfo {title} {Quantum computation in a hybrid array of molecules and rydberg atoms},\ }\href {https://doi.org/10.1103/PRXQuantum.3.030340} {\bibfield  {journal} {\bibinfo  {journal} {PRX Quantum}\ }\textbf {\bibinfo {volume} {3}},\ \bibinfo {pages} {030340} (\bibinfo {year} {2022})}\BibitemShut {NoStop}%
\bibitem [{\citenamefont {Zhang}\ \emph {et~al.}(2026)\citenamefont {Zhang}, \citenamefont {Murciano}, \citenamefont {Tantivasadakarn},\ and\ \citenamefont {Finkelstein}}]{Zhang2026HybridGate}%
  \BibitemOpen
  \bibfield  {author} {\bibinfo {author} {\bibfnamefont {C.}~\bibnamefont {Zhang}}, \bibinfo {author} {\bibfnamefont {S.}~\bibnamefont {Murciano}}, \bibinfo {author} {\bibfnamefont {N.}~\bibnamefont {Tantivasadakarn}},\ and\ \bibinfo {author} {\bibfnamefont {R.}~\bibnamefont {Finkelstein}},\ }\href {https://doi.org/10.48550/arXiv.2602.12909} {\bibinfo {title} {Quantum logic control and entanglement in hybrid atom-molecule arrays}} (\bibinfo {year} {2026}),\ \Eprint {https://arxiv.org/abs/2602.12909} {arXiv:2602.12909 [quant-ph]} \BibitemShut {NoStop}%
\bibitem [{\citenamefont {Guttridge}\ \emph {et~al.}(2023)\citenamefont {Guttridge}, \citenamefont {Ruttley}, \citenamefont {Baldock}, \citenamefont {Gonz\'alez-F\'erez}, \citenamefont {Sadeghpour}, \citenamefont {Adams},\ and\ \citenamefont {Cornish}}]{PhysRevLett.131.013401}%
  \BibitemOpen
  \bibfield  {author} {\bibinfo {author} {\bibfnamefont {A.}~\bibnamefont {Guttridge}}, \bibinfo {author} {\bibfnamefont {D.~K.}\ \bibnamefont {Ruttley}}, \bibinfo {author} {\bibfnamefont {A.~C.}\ \bibnamefont {Baldock}}, \bibinfo {author} {\bibfnamefont {R.}~\bibnamefont {Gonz\'alez-F\'erez}}, \bibinfo {author} {\bibfnamefont {H.~R.}\ \bibnamefont {Sadeghpour}}, \bibinfo {author} {\bibfnamefont {C.~S.}\ \bibnamefont {Adams}},\ and\ \bibinfo {author} {\bibfnamefont {S.~L.}\ \bibnamefont {Cornish}},\ }\bibfield  {title} {\bibinfo {title} {Observation of rydberg blockade due to the charge-dipole interaction between an atom and a polar molecule},\ }\href {https://doi.org/10.1103/PhysRevLett.131.013401} {\bibfield  {journal} {\bibinfo  {journal} {Phys. Rev. Lett.}\ }\textbf {\bibinfo {volume} {131}},\ \bibinfo {pages} {013401} (\bibinfo {year} {2023})}\BibitemShut {NoStop}%
\bibitem [{\citenamefont {Guttridge}\ \emph {et~al.}(2025)\citenamefont {Guttridge}, \citenamefont {Hepworth}, \citenamefont {Ruttley}, \citenamefont {Durst}, \citenamefont {Eiles},\ and\ \citenamefont {Cornish}}]{PhysRevLett.134.133401}%
  \BibitemOpen
  \bibfield  {author} {\bibinfo {author} {\bibfnamefont {A.}~\bibnamefont {Guttridge}}, \bibinfo {author} {\bibfnamefont {T.~R.}\ \bibnamefont {Hepworth}}, \bibinfo {author} {\bibfnamefont {D.~K.}\ \bibnamefont {Ruttley}}, \bibinfo {author} {\bibfnamefont {A.~A.~T.}\ \bibnamefont {Durst}}, \bibinfo {author} {\bibfnamefont {M.~T.}\ \bibnamefont {Eiles}},\ and\ \bibinfo {author} {\bibfnamefont {S.~L.}\ \bibnamefont {Cornish}},\ }\bibfield  {title} {\bibinfo {title} {Individual assembly of two-species rydberg molecules using optical tweezers},\ }\href {https://doi.org/10.1103/PhysRevLett.134.133401} {\bibfield  {journal} {\bibinfo  {journal} {Phys. Rev. Lett.}\ }\textbf {\bibinfo {volume} {134}},\ \bibinfo {pages} {133401} (\bibinfo {year} {2025})}\BibitemShut {NoStop}%
\bibitem [{\citenamefont {Wei}\ \emph {et~al.}(2026)\citenamefont {Wei}, \citenamefont {Artoni}, \citenamefont {La~Rocca}, \citenamefont {Wu},\ and\ \citenamefont {Shao}}]{7zjs-73qm}%
  \BibitemOpen
  \bibfield  {author} {\bibinfo {author} {\bibfnamefont {Y.}~\bibnamefont {Wei}}, \bibinfo {author} {\bibfnamefont {M.}~\bibnamefont {Artoni}}, \bibinfo {author} {\bibfnamefont {G.~C.}\ \bibnamefont {La~Rocca}}, \bibinfo {author} {\bibfnamefont {J.~H.}\ \bibnamefont {Wu}},\ and\ \bibinfo {author} {\bibfnamefont {X.~Q.}\ \bibnamefont {Shao}},\ }\bibfield  {title} {\bibinfo {title} {Enhancing ground-state interaction strength of neutral atoms via floquet stroboscopic dynamics},\ }\href {https://doi.org/10.1103/7zjs-73qm} {\bibfield  {journal} {\bibinfo  {journal} {Phys. Rev. A}\ }\textbf {\bibinfo {volume} {113}},\ \bibinfo {pages} {032812} (\bibinfo {year} {2026})}\BibitemShut {NoStop}%
\bibitem [{\citenamefont {Jackson}(1999)}]{Jackson1999}%
  \BibitemOpen
  \bibfield  {author} {\bibinfo {author} {\bibfnamefont {J.~D.}\ \bibnamefont {Jackson}},\ }\href@noop {} {\emph {\bibinfo {title} {Classical Electrodynamics}}},\ \bibinfo {edition} {3rd}\ ed.\ (\bibinfo  {publisher} {Wiley},\ \bibinfo {address} {New York},\ \bibinfo {year} {1999})\BibitemShut {NoStop}%
\bibitem [{\citenamefont {Saffman}\ \emph {et~al.}(2010)\citenamefont {Saffman}, \citenamefont {Walker},\ and\ \citenamefont {Mølmer}}]{Saffman2010}%
  \BibitemOpen
  \bibfield  {author} {\bibinfo {author} {\bibfnamefont {M.}~\bibnamefont {Saffman}}, \bibinfo {author} {\bibfnamefont {T.~G.}\ \bibnamefont {Walker}},\ and\ \bibinfo {author} {\bibfnamefont {K.}~\bibnamefont {Mølmer}},\ }\bibfield  {title} {\bibinfo {title} {Quantum information with {Rydberg} atoms},\ }\href {https://doi.org/10.1103/RevModPhys.82.2313} {\bibfield  {journal} {\bibinfo  {journal} {Reviews of Modern Physics}\ }\textbf {\bibinfo {volume} {82}},\ \bibinfo {pages} {2313} (\bibinfo {year} {2010})}\BibitemShut {NoStop}%
\end{thebibliography}%
\end{document}